\documentclass[11pt]{article}
\usepackage[english]{babel}
\usepackage{natbib}
\usepackage{rotating}
\usepackage{url}
\usepackage[utf8x]{inputenc}
\usepackage{amsmath,amssymb}
\usepackage{fancyhdr}
\usepackage[font={it}]{caption}
\usepackage{hyperref}
\usepackage{amsmath}
\usepackage{graphicx}
\usepackage{etoolbox}
\graphicspath{{images/}}
\usepackage{fancyhdr}
\usepackage[euler]{textgreek}
\usepackage{vmargin}
\setmarginsrb{2 cm}{1. cm}{2 cm}{1.5 cm}{1.5 cm}{.5 cm}{1.5 cm}{1.5 cm}

\title{Atmospheric characterization of terrestrial exoplanets in the mid-infrared: biosignatures, habitability \& diversity}								
\date{today's date}

\makeatletter
\let\thetitle\@title
\let\theauthor\@author
\let\thedate\@date
\makeatother

\def\aj{AJ}%
\def\apj{ApJ}%
\def\apjl{ApJ}%
\def\ao{Appl.~Opt.}%
\def\aap{A\&A}%
\def\mnras{MNRAS}%
\def\pasp{PASP}%
\def\nat{Nature}%

\begin{document}


\begin{titlepage}
	\centering
    \vspace*{0.5 cm}
\begin{center}    \textsc{\Large   White paper for the Voyage 2050 long-term plan \\in the ESA Science Programme}\\[1. cm]	\end{center}
	\textsc{\Large}\\[0.5 cm]				
	\rule{\linewidth}{0.5 mm} \\[0.4 cm]
	{ \LARGE\bf\thetitle}\\
	\rule{\linewidth}{0.5 mm} \\[1.5 cm]
	
            
		    \textsc{
			Prof. Sascha P. Quanz\\
			Swiss Federal Institute of Technology (ETH Zurich)\\
			Department of Physics\\
			Institute for Particle Physics and Astrophysics\\
			Exoplanets \& Habitability Group\\
            Wolfgang-Pauli-Strasse 27\\
            CH-8093 Zurich\\
            Switzerland\\
            Phone: +41 44 63 32830\\
            Mail: sascha.quanz@phys.ethz.ch\\}
           
    \vspace*{2. cm}
	
	\includegraphics[width=\linewidth]{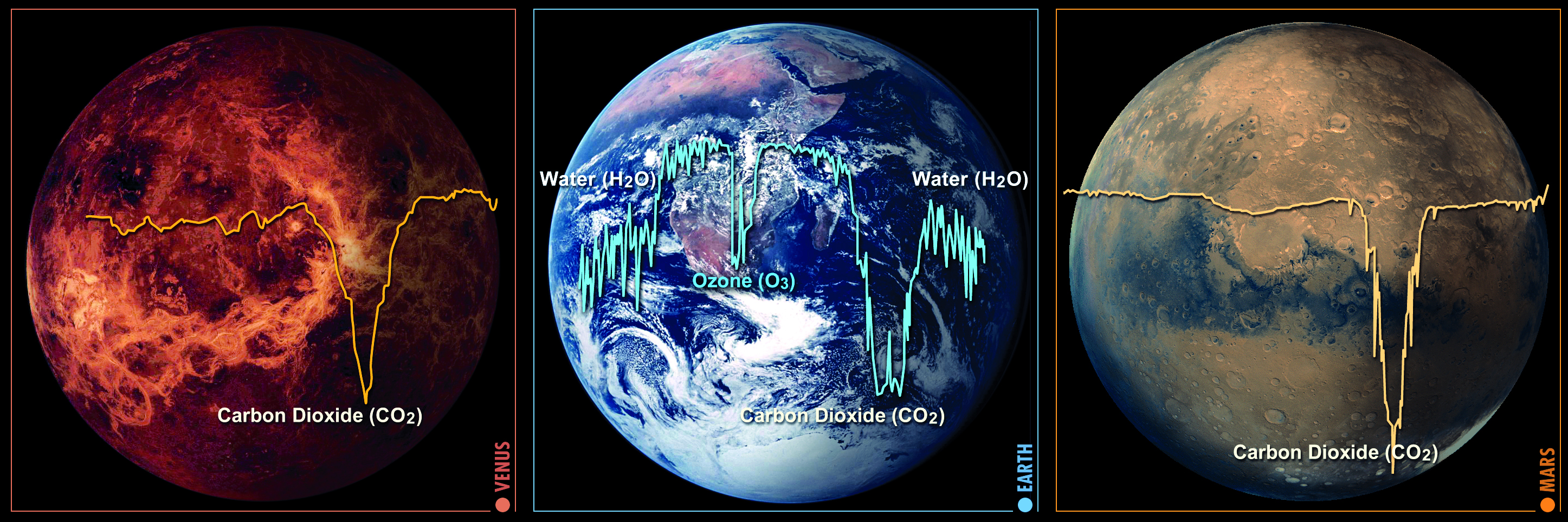}
    Thermal infrared emission spectra from different terrestrial planets. \\\footnotesize{(Image credit: ESA 2001; illustration by Medialab)}

\end{titlepage}



\renewcommand{\thesection}{\arabic{section}}

\section*{Prologue}
\thispagestyle{empty}
\vspace{1.5cm}
``To use Newton’s words, our efforts up till this moment have but turned over a pebble or shell here and there on the beach, with only a forlorn hope that under one of them was the gem we were seeking. Now we have the sieve, the minds, the hands, the time, and, particularly, the dedication to find those gems—no matter in which favorite hiding place the children of distant worlds have placed them.''\\\\
{\footnotesize Frank Drake and Dava Sobel, Is Anyone Out There? (1993)}
\vspace{1.cm}
\begin{figure}[h!]
\centering
    \includegraphics[width=\linewidth]{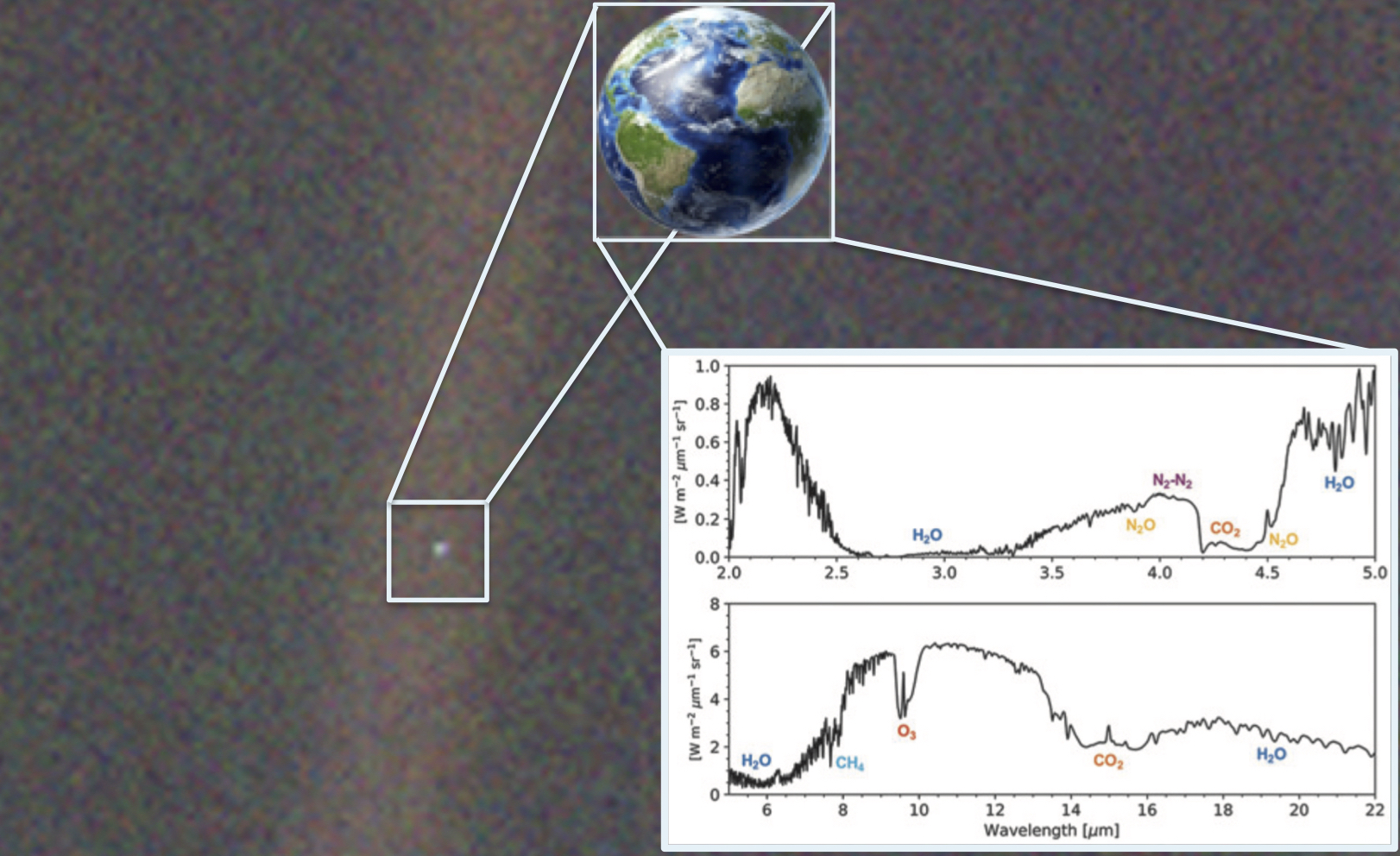}
    \label{fig:voyager}
\end{figure}

`Pale blue dot':\\Image of Earth taken on February 14, 1990, by the Voyager 1 space probe from a distance of about 40.5 AU. Spatially resolving the planet from its host star and analyzing its thermal emission spectrum in the $\sim$3--20 \textmu m range reveals Earth's -- thus far unique -- atmospheric fingerprints including the biosignature gases O$_3$, CH$_4$ and N$_2$O. \citep[Model spectrum from][]{schwieterman2018}
\newpage

\section{Scientific motivation and objectives}
\pagestyle{fancy}
\fancyhf{}
\fancyhead[CE,CO]{\small{Atmospheric characterization of terrestrial exoplanets in the MIR: biosignatures, habitability \& diversity}}
\fancyfoot[CE,CO]{\thepage}
\pagenumbering{arabic}
\subsection{Driving questions}
Exoplanet research is a focal point in modern astrophysics and one of the long-term primary objectives is the investigation of the atmospheric properties of dozens of small and terrestrial exoplanets. This is partially driven by the goal to search for and identify potentially habitable or even inhabited exoplanets. In addition, such a dataset is invaluable for investigating and understanding the diversity of planetary bodies. Exoplanet science is already omnipresent on the roadmaps of all major space agencies. However, none of the currently selected missions, neither in Europe nor in the US, will be able to deliver the above-mentioned comprehensive dataset of terrestrial exoplanet atmospheres as we will further detail below. The same is true for current and future ground-based observatories including the 30-40 m Extremely Large Telescopes (ELTs). Therefore, tackling a prime objective of exoplanet science and understanding how unique or common planets like our Earth are in our galactic neighborhood will require a new, dedicated approach.
Now is the right time to start investigating how a statistically relevant number of terrestrial exoplanet atmospheres can be analyzed and discuss how the guiding scientific objectives should be formulated. In fact, thanks to NASA's Kepler and TESS missions, ESA's upcoming PLATO mission, and ongoing and future radial velocity (RV) surveys from the ground, by 2030 we will have a robust statistical understanding of the occurrence rate of terrestrial exoplanets and their radius, mass and period distributions out to the habitable zone around main sequence stars, and we will have identified dozens of exoplanets in the immediate vicinity of the Sun including potentially habitable ones.  

The next logical step is to address the following questions:
\begin{itemize}
    \item [(Q1)] How many exoplanets exhibit (atmospheric) signatures of potential biological activity?
    \item [(Q2)] What fraction of terrestrial exoplanets 
    provide (surface) conditions so that liquid water and life as we know it \emph{could} in principle exist?
    \item [(Q3)] How diverse are (terrestrial) exoplanet atmospheres in their composition across a range of relevant parameters (e.g., planet mass and radius, host star spectral type, orbital period) and how does this compare to theories for planet and atmosphere formation and evolution?  
\end{itemize}
The sequence of these questions is deliberately chosen such as to go from the most specific (the search for biosignatures) to the most general (atmospheric diversity).  

\subsection{The mid-infrared opportunity}
While in-situ and/or fly-by measurements can in principle be carried out for Solar System bodies, this is not possible for exoplanets because of their distance. Instead, we have to rely on remote sensing techniques. These include investigations in reflected light (at optical and near-infrared (NIR) wavelengths), transmitted light (if the planet is transiting in front of its host star), or thermal emission (either through secondary eclipse, phase curve measurements or spatially resolved observations, all done at NIR to mid-infrared (MIR) wavelengths). {\bf We will argue in the following that spatially resolved observations in the MIR that aim to detect exoplanet thermal emission spectra are likely the most promising and powerful approach to address Q1-Q3 listed above and are hence the focus of this White Paper.} The scientific potential of studying (terrestrial) exoplanets in reflected light is discussed in the complementary White Paper by Snellen et al.  

\begin{figure}[h!]
\vspace{1.5cm}
    \centering
    \includegraphics[width=\linewidth]{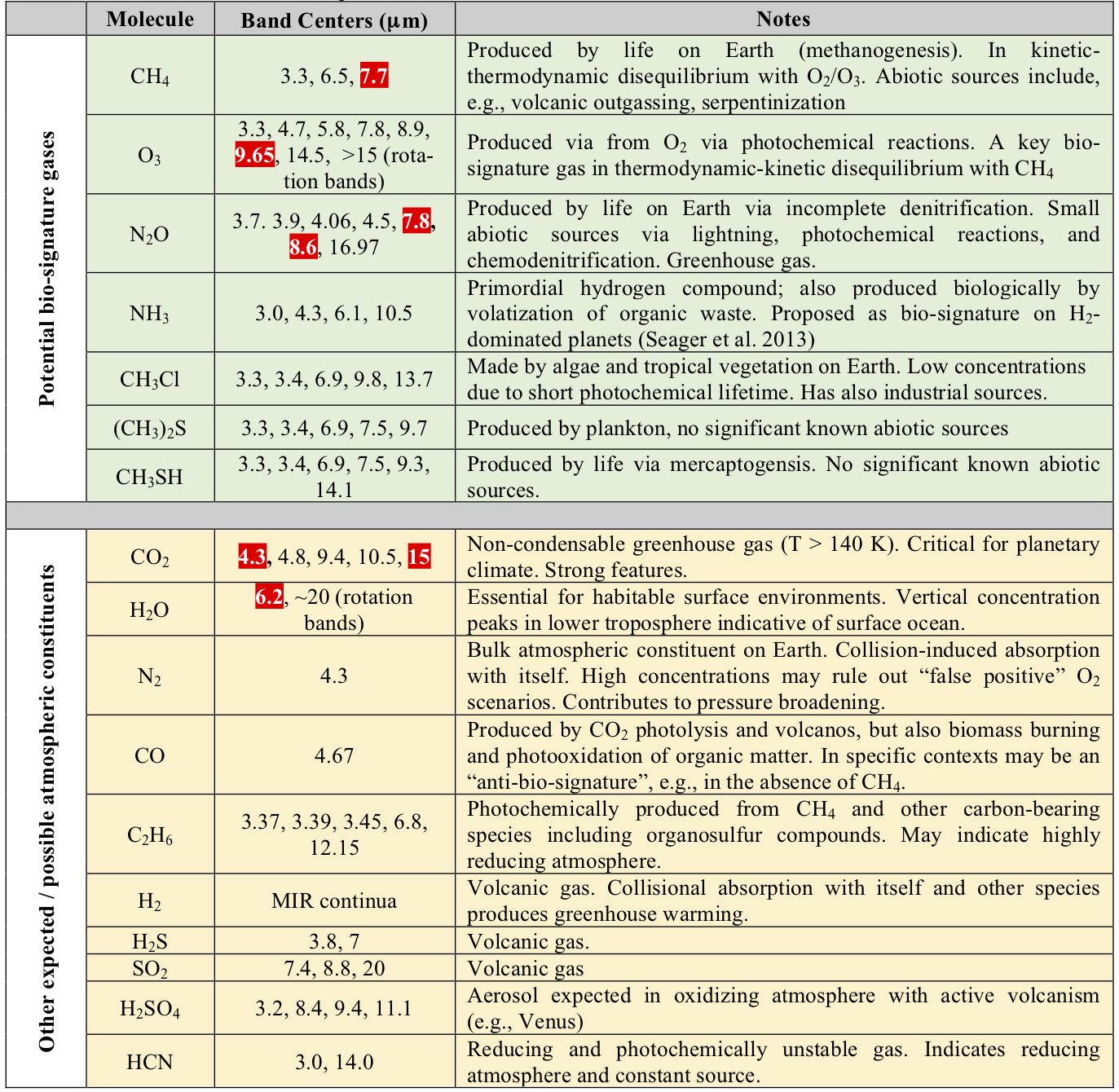}
    \caption{Molecules relevant to terrestrial planet characterization between 3-20 \textmu m (adapted from \cite{catling2018} and \cite{schwieterman2018}). Potential biosignature gases are listed in the top-half (green background) and other possible constituents in the lower half (yellow background). Strong bands in Earth's modern spectrum are highlighted in red. In particular the existence of significant CH$_4$ and N$_2$O bands is important to highlight as they have no strong counterparts at optical or NIR wavelengths. We also note the strong CO band around 4.67 \textmu m, which can serve as an ``anti-biosignature'' gas under certain circumstances.}
    \label{fig:molecules}
    \vspace{1.cm}
\end{figure}

For mature ($>$Gyr old) planets orbiting in the inner few AU around their host stars the energy budget of their atmospheres is typically dominated by the absorption and re-radiation of stellar energy. The temperature structure of the atmosphere, i.e., the temperature as a function of height or pressure, is a key diagnostic and a driver of chemistry and climate. An emission spectrum encodes information about this temperature structure as well as the re-radiated luminosity of the planet, which -- in combination with the observed effective temperature -- strongly constrains the planet radius. Taking the Earth as reference, certain atmospheric windows in the MIR may even allow a direct measurement of the surface temperature of a terrestrial exoplanet \citep[e.g.,][]{desmarais2002}. Furthermore, the MIR wavelength regime offers an unparalleled diagnostic potential to determine the atmospheric composition as multiple major molecules required to explore planetary conditions present strong absorption bands in the MIR. Thermal emission observations are also less influenced by (though not insensitive to) the presence of clouds \citep[e.g.,][]{kitzmann2011}. Mitigating the role of uncertain cloud properties is imperative to our understanding of atmospheric composition. In particular for the detection of biosignature gases and chemical disequilibrium,  the MIR is an ideal spectral regime \citep[e.g.,][]{desmarais2002,schwieterman2018}. The MIR includes absorption bands from ozone (O$_3$) and methane (CH$_4$) and the presence of both molecules in the Earth atmosphere, which unless continuously replenished would quickly react with each other, can only be explained by biological activity. Furthermore, nitrous oxide (N$_2$O), another potential biosignature present in the Earth atmosphere, has strong bands in the MIR, but, similar to methane, no equally strong bands at optical or NIR wavelengths. Molecular oxygen (O$_2$) can be detected in the optical regime (around 760-765 nm) and not in the MIR, but interpreting oxygen as a potential biomarker requires additional contextual information \citep{meadows2018}. The unique richness of the MIR spectral range in general, but in particular in the context of potential biosignature gases, is summarized in Figure~\ref{fig:molecules}.
\begin{figure}[h!]
    \centering
    \includegraphics[width=0.97\linewidth]{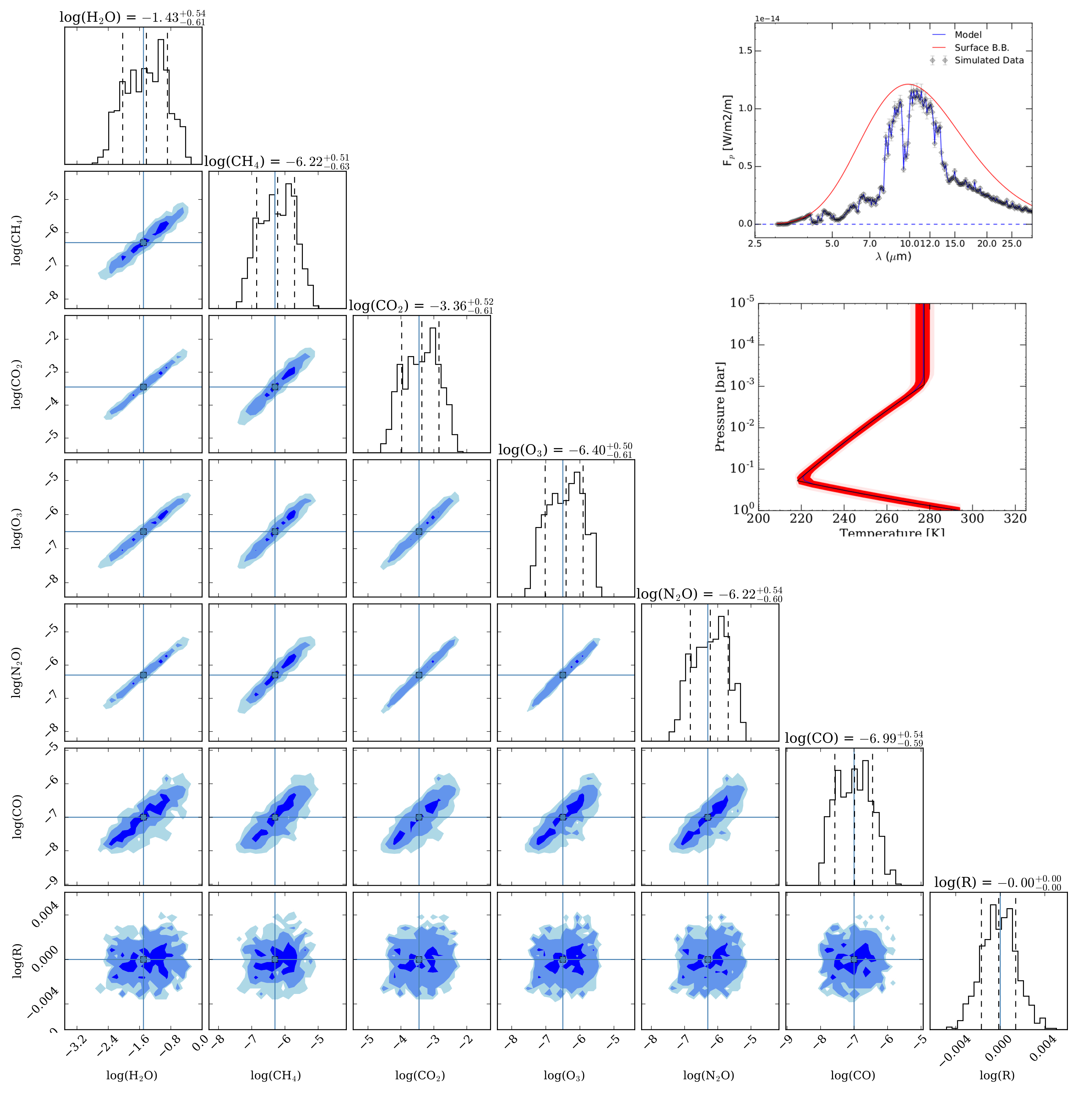}
    \caption{Results of a retrieval study of an Earth-twin atmosphere observed over a wavelength range of 3-30 \textmu m with $R=100$ and $SNR=20$. The inlay in the top right corner shows the model in blue, the simulated observational data in black and a black-body curve representing surface emission in red. Below, the corresponding pressure-temperature profile is plotted with the model again in blue, the best-fit retrieved profile in black and the red band indicating the 68\% confidence range. The corner plot shows the marginalized distributions for the abundance of major molecules detected in the atmosphere as well as for the radius of the planet, all of which were free parameters in the retrieval (with flat priors). All parameters are well constrained. We note that also the planet's mass was a free parameter in the retrieval analysis (with a flat prior confined to the range 0.1-10 M$_\oplus$). As expected, it  could not be accurately estimated from the spectrum alone and complementary RV measurements or an empirically calibrated mass-radius relationship \citep[e.g.,][]{chen2017} would be  needed to provide constraints.}
    \label{fig:retrieval1}
\end{figure}

The following three sub-sections provide more quantitative information regarding the driving questions listed in section 1.1 above. Section 1.2.1 deals with the admittedly  hypothetical case of observing an Earth-twin and searching for biosignatures (Q1). In section 1.2.2 we discuss how one could statistically address the fundamental question how common or rare terrestrial exoplanets with Earth-like (surface) conditions are (Q2). Finally, in section 1.2.3 we show how many exoplanets covering a broad range of sizes and orbital periods could be detected by a MIR exoplanet imaging mission in order to investigate atmospheric diversity in the most general sense (Q3).

\subsubsection{Atmospheric modeling and retrieval of key parameters for an Earth-twin (Q1)}
We carried out a spectral retrieval study \citep[see, e.g., ][for a recent review on atmospheric retrieval for exoplanets]{madhu2018}, where Earth's emission spectrum was modeled over a certain wavelength range assuming a certain spectral resolution ($R$) and signal-to-noise ($SNR$) per resolution element \citep[cf.][]{vonparis2013,leger2019}. Taking such a model spectrum, i.e., a simulated observation of an Earth-twin, as input, the retrieval framework allows us to derive posterior distributions for key atmospheric parameters including the pressure-temperature profile, abundances of key atmospheric constituents as well as the radius of the planet. We acknowledge that the choice of an Earth-twin exoplanet is a special case and it seems unlikely that the formation and subsequent evolution processes that led to the Earth's present day atmosphere took also place on another nearby exoplanet. However, we argue that any mission that aims to detect potential biosignatures in an exoplanet's atmosphere should be able to identify such  signatures in our Earth's atmosphere, the only planet we know to harbor life. Hence, taking an Earth-twin as starting point, as is also frequently done in other publications \citep[e.g.,][]{kitzmann2010,kitzmann2011,gebauer2017}, appears justifiable. 

Our results from one specific simulation are shown in Figure~\ref{fig:retrieval1}. Here, we assumed a wavelength range of 3-30 \textmu m, $R=100$ and $SNR = 20$; however, the results would hardly change if the wavelength range were limited to 3-20 \textmu m. The values for $R$ and $SNR$ were chosen  to quantitatively compare the results to those derived in a published retrieval study by \cite{feng2018} who investigated the diagnostic potential of reflected light observations of an Earth-twin in the 0.4-1.0 \textmu m range\footnote{\citet{feng2018} presented different scenarios and for the comparison we selected the case where they assumed $R=140$ and $SNR=20$ (at 550 nm), i.e., comparable to the values we assume.}. 

Figure~\ref{fig:retrieval1} clearly demonstrates that if one were able to obtain a high-quality MIR spectrum of an Earth twin, the atmospheric composition, but also surface pressure, surface temperature and planetary radius could be constrained with high precision and accuracy.

\begin{figure}[t!]
    \centering
     \includegraphics[width=0.85\linewidth]{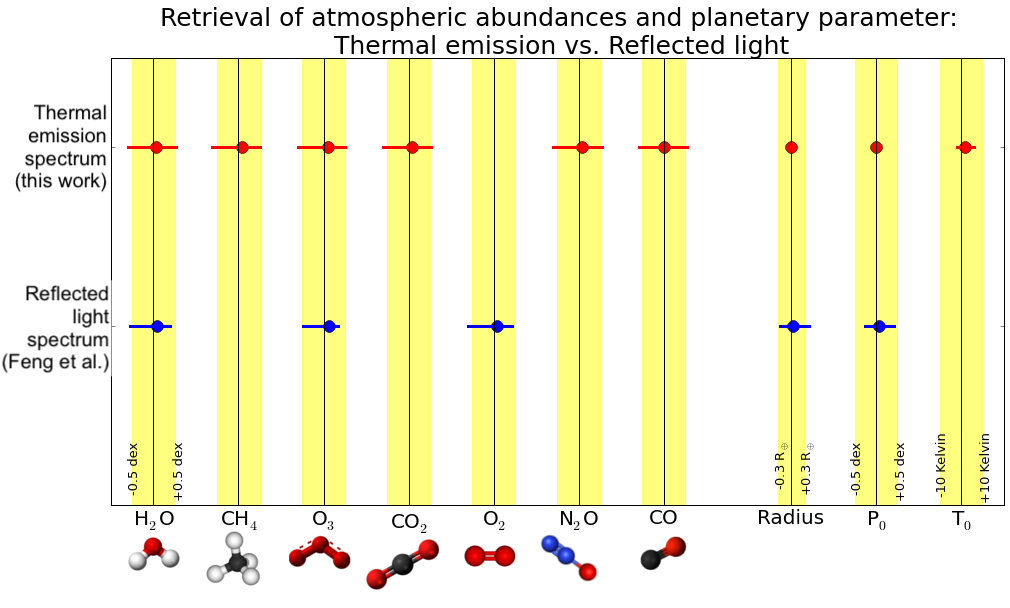}
    \caption{Comparing the retrieval results for the simulated thermal emission spectrum shown in Figure~\ref{fig:retrieval1} and those found in \citet{feng2018} for reflected light to the input values used for the different parameters in the atmospheric models. The black vertical lines represent the `true' value for each parameter, the yellow areas indicate the $\pm$0.5 dex range for the atmospheric constituents and for the surface pressure, the $\pm$0.3 R$_{\oplus}$ range for the planet radius and the $\pm$10 K range for the surface temperature. The red and blue data points show the location of the retrieved best-fit parameters, for thermal emission and reflected light, respectively, and the error bars indicate their 68\% confidence range.}
    \label{fig:retrieval2}
\end{figure}

This is further illustrated in Figure~\ref{fig:retrieval2} where the best-fit values as well as the 68\% confidence range for the retrieved parameters are plotted (red data points) and compared to the `ground-truth', i.e., the input value in the simulated spectrum (black lines). In all cases, the input value is well within the  68\% confidence range and this confidence range typically corresponds to a factor of 3 ($\sim$0.5 dex) uncertainty for the molecular abundances. Furthermore, the radius is constrained to much better than a few percent, the surface pressure to 0.1 dex and the surface temperature to better than 5 K. We emphasize that the radius was determined solely from the emission spectrum and no additional information from, e.g., transit observations, were required. {\bf With such a dataset in hand, assessing the atmospheric conditions -- and potentially the surface conditions -- of this planet would easily be feasible. The existence of atmospheric biosignatures could be detected with high confidence.}

In addition to the results from our retrieval study, Figure~\ref{fig:retrieval2} also shows the same analysis using the results found by \citet{feng2018} for reflected light (blue data points) instead of thermal emission. 
{\bf The key take-away messages from this comparison are the following:}
\begin{description}
    \item (a) The atmospheric abundances that can be constrained using thermal emission are retrieved with similar accuracy and precision as those accessible in reflected light.
    \item (b) The biosignature gases CH$_4$ and N$_2$O are \emph{only} accessible in the thermal emission spectrum as is the anti-biosignature gas CO.
    \item (c) O$_2$ is only accessible in reflected light, but in the MIR  emission spectrum O$_3$, which is an atmospheric by-product of O$_2$, serves as a robust proxy for the existence of oxygen.
    \item (d) The constraints on the planetary radius are much stronger in thermal emission (uncertainties are of order 5\%) compared to reflected light (uncertainties are of order 30\%) as there is a degeneracy between the planet's albedo and its radius in reflected light.
    \item (e) While surface temperature and pressure can be well constrained with thermal emission spectra, reflected light does not provide immediate information about surface temperature and constraints on pressure are weaker. 
\end{description}

{\bf This comparison demonstrates the enormous characterization potential contained in the thermal emission spectra of terrestrial exoplanets.}  Not included in this analysis is the possibility of detecting N$_2$-N$_2$ collision-induced absorption features in the MIR \citep[see, Figure~\ref{fig:molecules};][]{schwieterman2015}, which could help constrain the absolute abundances of the different components in N$_2$-dominated atmospheres. For reflected light studies at optical wavelengths surface features such as `glint' from water oceans could be detected under certain conditions \citep[e.g.,][]{LustigYaeger2018} and also the so-called `red-edge', the specific spectral reflectance of chlorophyll in plants, is observable -- at least in case of the Earth \citep[e.g.,][]{seager2005}.


\subsubsection{Statistical significance and possible null-result (Q2)}
Already the detection of a single exoplanet spectrum with clear indications of biologically induced disequilibrium chemistry would be a breakthrough result that warrants special care in interpreting  \citep{catling2018}. However, one must be prepared for a null-result and a future mission should be able to address more general -- and scientifically equally important -- questions related to the population of (terrestrial) exoplanets, their atmospheres and climates. Q2 and Q3 listed above capture such questions and require a sample of exoplanets to be investigated. In particular for Q2 the sample size should be defined in such a way that the null-result, i.e., that none of the planets that are characterized turns out to provide conditions similar to those on Earth, is a major scientific result and robustly quantifies the rareness of ``Earth-like'' planets. A possible approach is to re-formulate this question in the following hypothesis: 

\vspace{0.2cm}
\emph{``The fraction of terrestrial exoplanets that reside in the empirical habitable zone around their host star \emph{and} provide conditions for liquid water to exist is $H_0$.''}
\vspace{0.2cm}

We emphasize that $H_0$ is \emph{not} identical to the commonly used $\eta_\oplus$ parameter that quantifies the fraction of stars that harbor terrestrial exoplanets in their habitable zone and $H_0$ is also ignorant of the potential existence of biosignatures in the atmospheres. In the context of the hypothesis above we define ``terrestrial exoplanets" as planets with a radius $R_{\rm planet}$ with $0.5\,R_{\oplus}\leq R_{\rm planet} \leq 1.5\, R_{\oplus}$ and the ``empirical habitable zone" as the separation range where the incoming stellar insolation $S$ is $0.35 \,S_\oplus \leq S \leq 1.75\, S_\oplus$ with $S_\oplus$ being the solar constant, i.e., the average insolation received at Earth \citep[cf.][]{kaltenegger2017}. The small end of the radius range is defined by the size of Mars, which is assumed to be the minimum planet size/mass that can retain an atmosphere, and the large end by the transition between rocky and gas dominated planets \citep{rogers2015,wolfgang2016,chen2017}. The insolation range is defined by the so-called ``Early Mars limit'' at the outer edge and the ``Recent Venus limit'' at the inner edge \citep{kopparapu2013}. We acknowledge that the concept of ``(exoplanet) habitability" is complex \citep[e.g.,][]{leconte2013} and extends beyond the radius and insolation ranges considered here \citep[e.g.,][]{zsom2013,seager2013}. However, similar to the case of the Earth-twin considered in section 1.2.1, we think that our choices are a reasonable and justifiable starting point that can be modified going-forward as our understanding of ``habitability" and exoplanet properties further progresses.

For the radius and insolation ranges considered here, two planets in the Solar System, Earth and Mars, qualify. Hence, the fraction of terrestrial exoplanets that reside in the empirical habitable zone in the Solar System \emph{and} allow for the existence of liquid water is one (Earth) out of two, i.e., $H_0=50\%$. 

\begin{figure}[h!]
    \centering
    \includegraphics[width=0.9\linewidth]{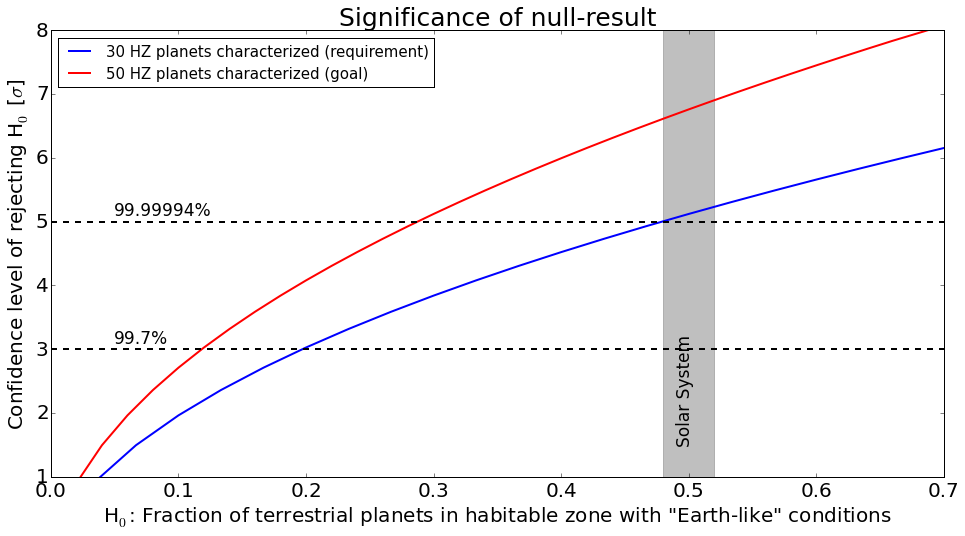}
    \caption{The statistical power of a null-result: in case 30 (blue curve) or 50 (red curve) exoplanets with radii between 0.5 and 1.5 R$_{\oplus}$ and receiving between 0.35 and 1.7 times the insolation of the Earth are investigated with high-quality thermal emission spectra and not a single one is found to support conditions that allow for the existence of liquid water, then the null-hypothesis -- shown on the x-axis -- can be rejected with the significance shown on the y-axis. In the Solar System, one out of two planets within the empirical habitable zone provides (surface) conditions for liquid water to exist; hence, H$_0=50\%$ for the Solar System.}
    \label{fig:null_result}
\end{figure}

In Figure~\ref{fig:null_result} we show the significance that a certain value for $H_0$ can be rejected in case 30 (blue line) or 50 (red line) planets were observed and none of them provides favorable conditions. These numbers are based on Poisson statistics and assume that the occurrence of terrestrial exoplanets orbiting in the habitable zones of different stars can be treated as statistically uncorrelated events. It shows that, if 30 planets are observed, $H_0=20\%$ and $H_0=50\%$ can be rejected with $\approx$3$\sigma$ and $\approx$5$\sigma$, respectively. For 50 planets, $H_0=10\%$ and $H_0=30\%$ can be rejected with the same confidence levels. These results suggest that, in case of a null-result, several tens of planets would be required in order to derive statistically significant limits on the rareness of  ``Earth-like'' planets. 

To date we do not yet have a large enough sample of exoplanets detected that fulfill the criteria used in the analysis (cf. section 4.2). However, steady progress is being made by ongoing surveys and missions to increase the number of relevant planets. Alternatively, a mission that can address Q1 and Q2 and the objectives formulated above could be split in two phases: (1) a ``search phase", aiming at quickly detecting a sufficient number of planets in the above-mentioned radius and insolation range; and (2) a ``characterization phase", where a sub-set of the detected planets would be re-observed and investigated with high SNR  in sufficient detail. To make the search phase time-efficient, a  broad-band photometry mode, e.g., by collapsing the MIR spectra over certain wavelength ranges to increase the SNR, could be applied. As we will detail below, it is important to mention that during the search phase many more planets, with properties outside the parameter range defined above, would be detected ``for free''. It turns out that if the occurrence rates of exoplanets (including their radius and period distributions), as found by NASA's Kepler mission, are applicable to exoplanets orbiting stars in the vicinity of the Sun, several hundred planets may be detected.

\subsubsection{Atmospheric diversity and total planet yield (Q3)}
Similar to the diversity in planet radii and orbital periods, as revealed by NASA's Kepler mission, we can expect a great diversity in atmospheric properties of (terrestrial) exoplanets \citep[e.g.,][]{leconte2015}. It is hence important to understand how many exoplanets in general, i.e., over a large region in the radius vs. stellar insolation parameter space, could be detected, e.g., during the search phase of a MIR exoplanet imaging mission. The detection of a large sample would enhance the value of the mission and the potential science legacy by enabling the exploration of (unbiased) planetary system architectures and the constraints they put on planet formation theory and atmospheric models. At the same time, the feasibility of detecting tens of ``Earth-like'' exoplanets as defined above in section 1.2.2 needs to be investigated. 
\begin{figure}[b!]
    \centering
\includegraphics[width=0.72\linewidth]{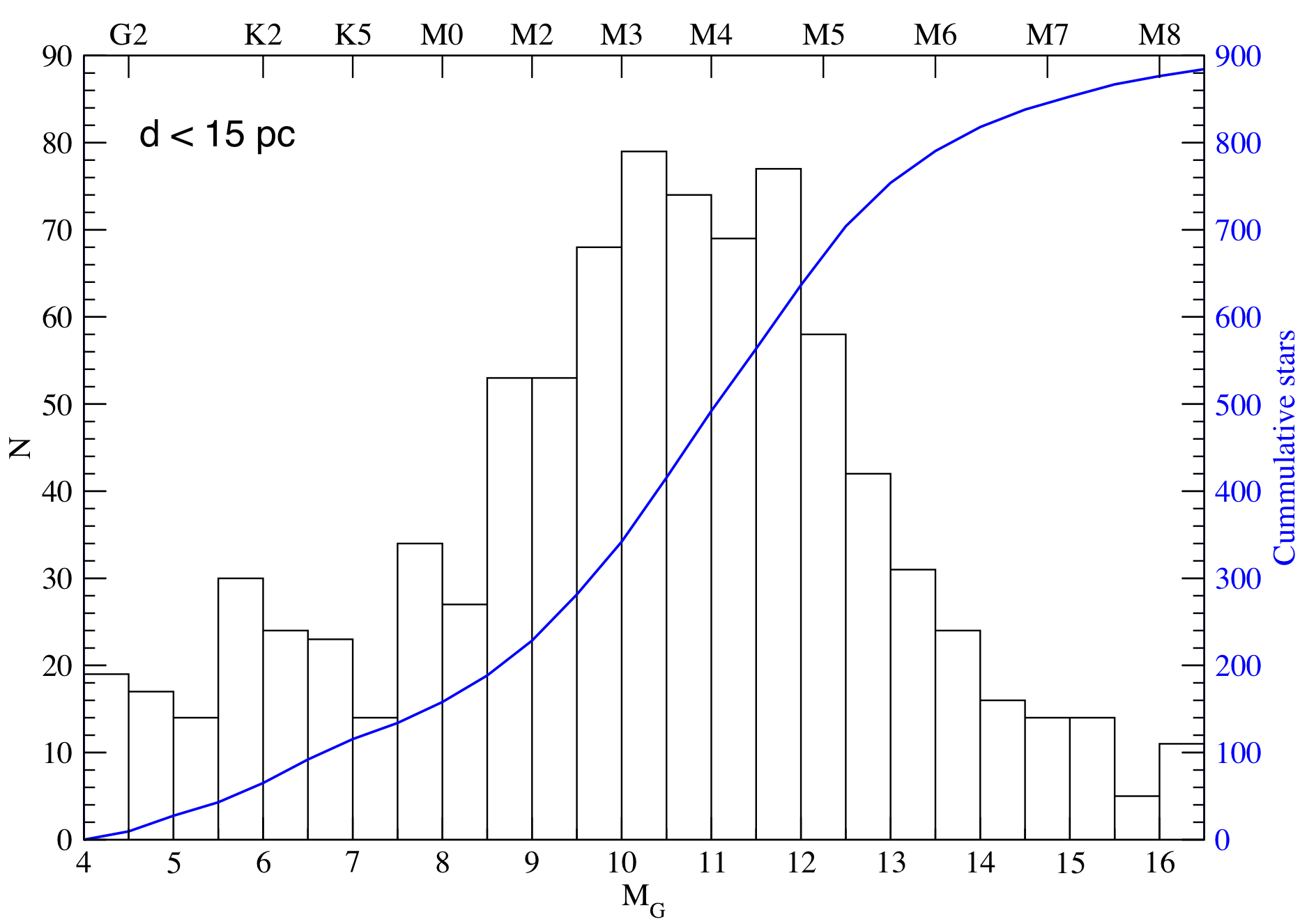}
    \caption{All known GKM main-sequence stars within 15 pc from the Sun as a function of their apparent G magnitude (GAIA filter). The left y-axis shows the number per bin in the black histogram, the right axis the cumulative number indicated by the blue line ($\sim$900 objects in total). The axis on the top shows the approximate location of different spectral types relative to G magnitudes. In addition to the stars shown here, there are 21 F-type stars known within 15 pc. 
    } 
    \label{fig:stars}
\end{figure}

We hence updated Monte Carlo simulations that were first presented in \citet{kammerer2018} and \citet{quanz2018} to quantify the exoplanet yield during the search phase mentioned above. The technical specifications used in these simulations are based on earlier concept studies for a space-based MIR interferometry mission \citep{cockell2009}, but updated with more recent estimates for sensitivity limits similar to those of the MIRI instrument on the James Webb Space Telescope (JWST). For more details on sensitivity and spatial resolution requirements we refer to section 2.2. For the underlying planet population around FGK stars, in terms of planet occurrence rate and radius and period distributions, we used the statistics published by NASA's working group SAG13\footnote{\url{https://exoplanets.nasa.gov/exep/exopag/sag/\#sag13}}. These statistics were also used for recent studies for reflected light missions \citep{kopparapu2018}. For M stars the statistics from \citet{dressing2015} were used. In general, Kepler and other ongoing missions and projects have impressively demonstrated that planetary systems are ubiquitous, including exoplanets close to or in the habitable zone \citep[e.g.,][]{burke2015,gillon2017,escude2016,hsu2019}. We now know that planets with sizes in the Earth and Super-Earth regime populate nearly every star and that systems consisting of multiple planets are very abundant \citep[e.g.,][]{weiss2018}.

The stellar sample we used in our simulations consisted of 320 F, G, K and M stars all within 20 pc from the Sun \citep{crossfield2013,kammerer2018}\footnote{Only 3 stars in the sample are of spectral type M6; all other stars have earlier spectral types.}. This is only a small subset of all stars within this distance limit (see, Figure~\ref{fig:stars}) and refining the possible target sample is subject of ongoing work. One important piece of information that needs to be carefully considered is the number of binary star systems that can in principle be included. Stellar binarity, over a certain range of separations, has been shown to have a negative impact on the occurrence of small planets \citep{kraus2016}. A first rough estimate would suggest that no more than 20-30\% of the stars shown in Figure~\ref{fig:stars} should be eliminated, leaving more than 600 potential targets within only 15 pc.

 For each target star that we consider in our simulations, 5000 planetary systems were created with properties randomly drawn from the distributions mentioned above and randomly oriented orbits and planetary positions thereon. Based on their apparent separation and estimated flux levels (from randomly drawn Bond albedos and assuming black body emission) we count all planets that are detectable according to the technical specifications (see section 2.2). We assume that the search phase could be carried out simultaneously in broad spectral bands centered at 5.6, 10, and 15 \textmu m (e.g., by collapsing the observed spectra around these wavelength ranges)
\begin{figure}[b!]
    \centering
    \includegraphics[width=0.49\linewidth]{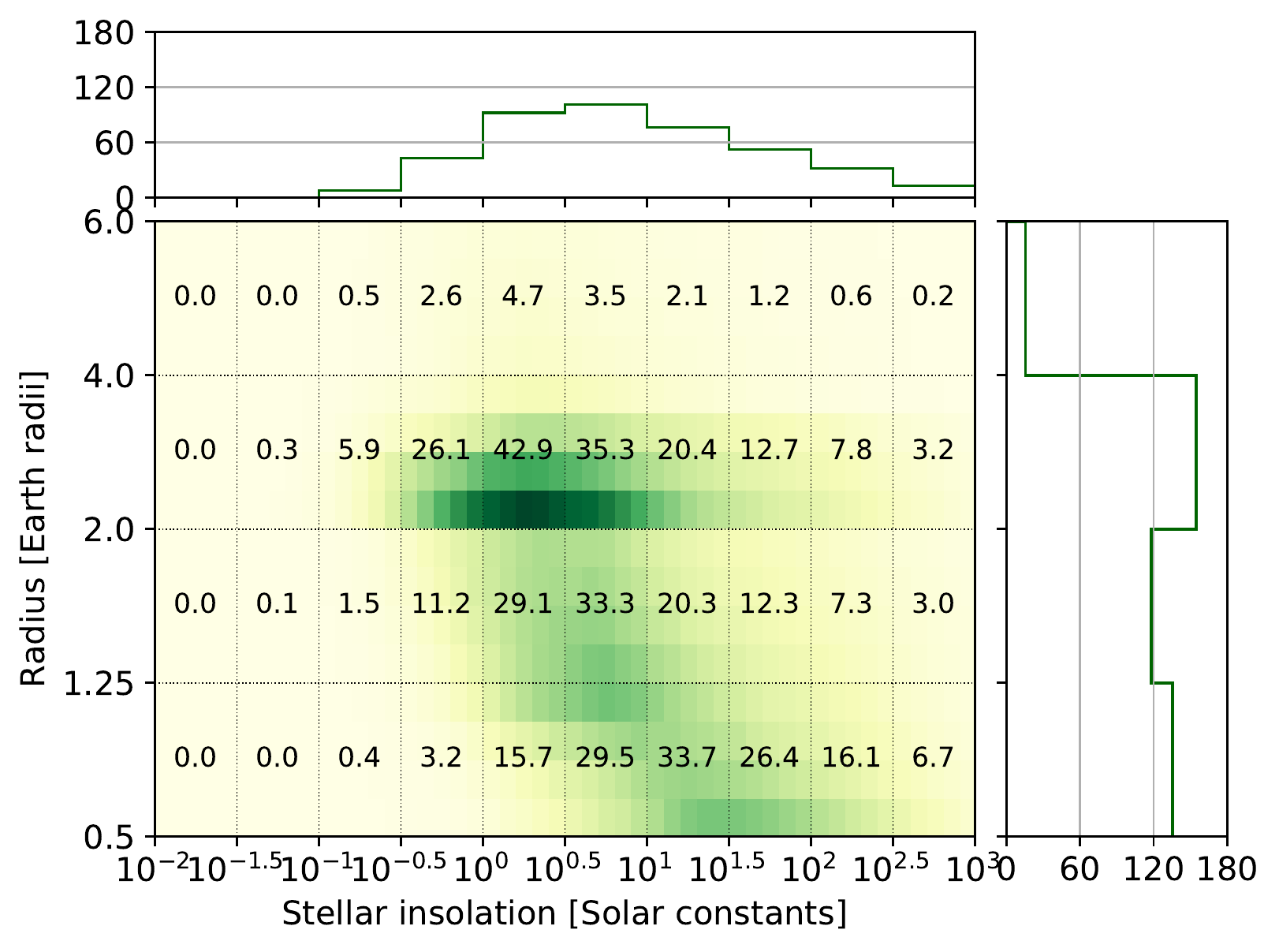}
    \includegraphics[width=0.49\linewidth]{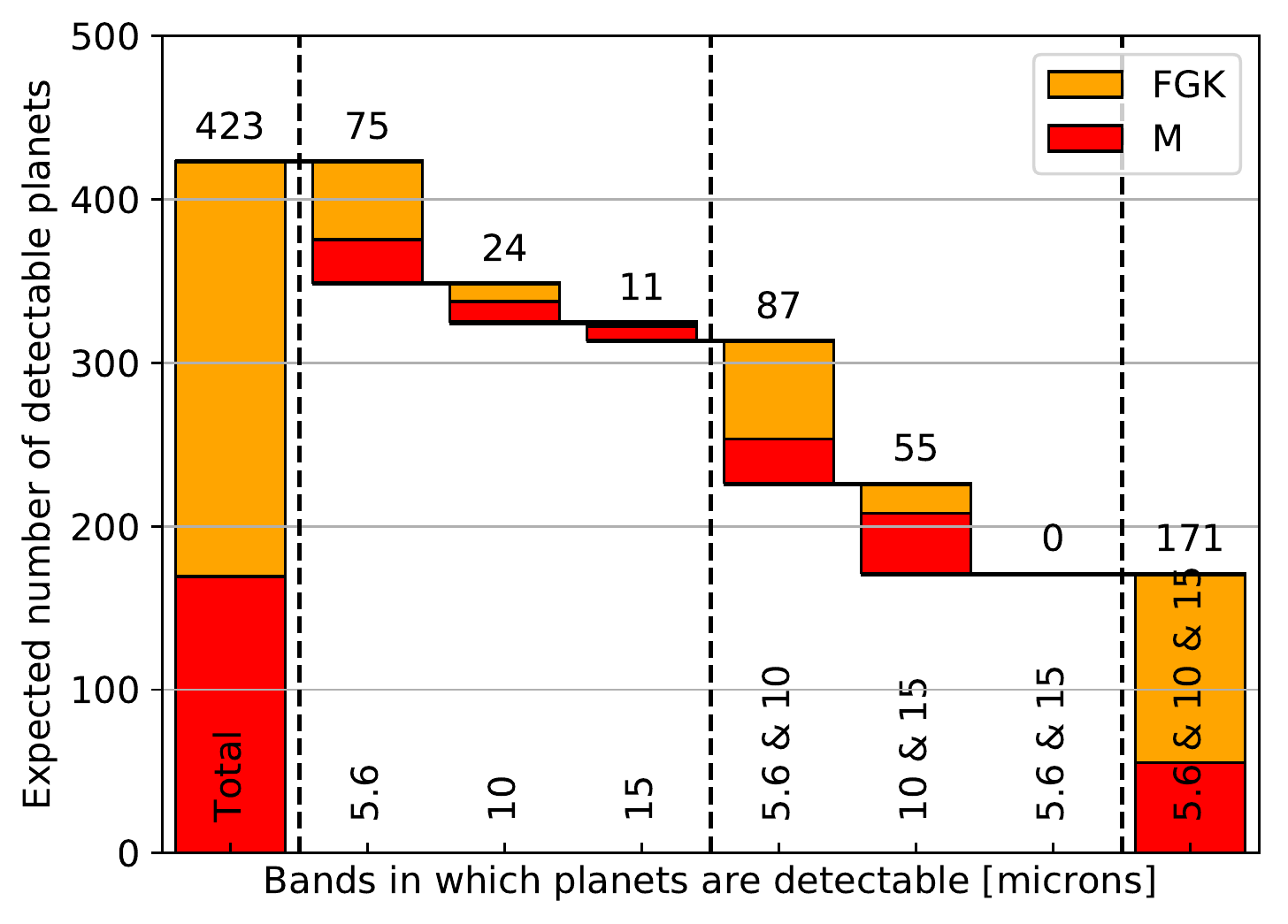}
    \caption{Estimated exoplanet yield in a hypothetical 3-year search phase for a space-based MIR exoplanet imaging mission targeting 320 stars within 20 pc and assuming planet statistics as derived by NASA's Kepler mission \citep[cf.][]{kammerer2018}. Left: Number of expected exoplanet detections per bin in the radius vs. stellar insolation plane. Right: The integrated number of expected exoplanets (423) broken down into sub-sets that are either detected in only one, two, or all three of the assumed bands centered at 5.6, 10, and 15 \textmu m. Planets detected around FGK stars are shown in orange, planets around M stars in red bars.}
    \label{fig:yield}
\end{figure}
and that all 320 stars are observed for an equal amount of time (35000 s). In total this amounts to less than 0.5 years and hence it seems realistic that even if overheads (e.g., slewing from one star to the next) and additional noise sources (e.g., stellar leakage or extra-zodiacal light) are considered, a potential search phase would not exceed 3 years. Figure~\ref{fig:yield} shows the results: {\bf the total number of detectable planets exceeds 400} and these planets cover a broad range of radii (0.5-6.0\,R$_{\oplus}$) and stellar insolation levels ($\approx$0.1-1000\,S$_{\oplus}$). In fact, as shown in the left panel of Figure~\ref{fig:yield}, in each of the radius bins covering 0.5-1.25, 1.25-2.0 and 2.0-4.0 R$_{\oplus}$ more than 120 exoplanets should be detectable. {\bf Such a database would be an excellent starting point to address Q3 listed above, the diversity of planetary atmospheres. Furthermore, among those $>$400 exoplanets, the number of detectable terrestrial planets located in the empirical habitable zone, as defined in section 1.2.2, is $\sim$30, which is the minimum number of planets needed for a meaningful statistical interpretation of a null-result (cf. Figure~\ref{fig:null_result}).} This number can be further increased by optimizing the distribution of observing time across the stellar target list in order to maximize the number of detectable terrestrial planets located in the empirical habitable zone \citep[cf.][]{stark2014}. Further optimization potential lies in the selection of the stellar target sample. 


\section{The need for a large space mission}
In the previous sections we already alluded to a possible mission scenario that would enable the science described above: a MIR exoplanet mission that would consist of (1) a search phase to detect a large sample of exoplanets in certain broad spectral bands, and (2) a follow-up characterization phase to investigate the (atmospheric) properties of a sub-set of these exoplanets with high-fidelity MIR low-resolution spectroscopy over the 3--20 \textmu m range. In the following we argue that this can only be achieved from space in the context of a large mission.
\subsection{Why space?}
Detecting thermal emission from celestial bodies between 3--20 $\mu$m wavelength from the ground is severely hampered by the thermal background emission caused by the Earth atmosphere \citep[$>$100 Jy/arcsec$^2$ for wavelengths $\gtrsim$8 \textmu m on Cerro Paranal, ESO's site for the Very Large Telescope (VLT);][]{absil2006}, the telescope's primary mirror and any additional non-cryogenic optical component in the light-path. For comparison, the Earth's emission spectrum peaks around 10~\textmu m (Figure~\ref{fig:retrieval1}), but seen from a distance of 10 pc our planet emits only $\sim$0.4~\textmu Jy at these wavelength ranges \citep{desmarais2002}. Consequently, even with the upcoming 30-40 m ELTs, only a handful of terrestrial exoplanets around the very nearest stars will be detectable at 10~\textmu m wavelength within a reasonable amount of observing time \citep{quanz2015}. Furthermore, the science described above requires a continuous wavelength coverage to measure the luminosity of the exoplanets and search for various atmospheric key components. In particular the water band at 6.2~\textmu m, and methane (CH$_4$) and nitrous oxide (N$_2$O) at 7.7 and 7.8~\textmu m, respectively, are not detectable from the ground. The two latter molecules are, however, important bio-signatures (Figure~\ref{fig:molecules}) that are key for addressing Q1. In conclusion, achieving the science goals of this White Paper requires going to space. 

\subsection{Why L-class?}
The direct detection of dozens of (small) exoplanets around nearby stars in the MIR  wavelength regime requires both high spatial resolution and high sensitivity. An Earth-like planet seen at 10 pc has an apparent separation of only 0.1$''$ from a solar-type host star and planets orbiting around M-stars must be even closer to their stars in order to be located in the habitable zone. Even JWST with its 6.5 m primary mirror does not provide sufficient spatial resolution ($\sim$0.45$''$ at 10~\textmu m). Furthermore, the low flux levels from terrestrial exoplanets (see above the example of Earth) set some important constraints on the collecting area of the telescope. {\bf To our knowledge, the only way to achieve both sufficient spatial resolution and sensitivity is a mid-infrared nulling interferometer}. It would consist of several spacecraft (`collector telescopes') that together provide enough collecting area, but are sufficiently far separated from each other so that the baselines between them provide the required spatial resolution. The beam combination would be done in a separate spacecraft where light from the central star interferes destructively (`nulling') as to provide sufficient contrast to detect the much fainter signal from the planets (typically of order 10$^{-7}$--10$^{-6}$ fainter at 10~\textmu m.). 

For the exoplanet yield estimate in section 1.2.3 we assumed a nulling baseline between the collector telescopes of up to $\sim$170 m. As a starting point for the achievable sensitivity limits we took those from JWST/MIRI \citep{glasse2015}, but reduced the overall instrument throughput by a factor of 3.5 \citep[an interferometer will likely have a lower throughput than a single-dish instrument, cf.][]{kammerer2018}. This means, however, that implicitly a collecting area similar to the effective aperture size of JWST was assumed. In case of 4 collector telescopes this translates into individual aperture sizes of the order 2.5 m in diameter (note: the primary mirror of the Hubble Space Telescope is 2.4 m and that of the Herschel Space Observatory was 3.5 m). The smaller the combined aperture size of the collector telescopes (or the smaller the number of collector telescopes), the longer the mission search phase to detect a large sample of exoplanets and the longer also the time needed to do high-SNR follow-up observations for in-depth atmospheric characterization. Indeed, a 2-telescope interferometer with apertures of a few tens of centimeter may only be able to search for planets around a few nearby stars \citep{defrere2018}. A large exoplanet sample that allows for a robust statistical analysis as needed to address Q2 and Q3 would no longer be feasible. 

This suggest that, very likely, a mission designed to address the science objectives described above can only be implemented in the framework of an ESA  L-class mission. However, given the ambitious science goals and unique scientific capabilities of such a mission delivering truly ground-breaking results, other international partners might be interested in a joint effort (see also below).    


\subsection{Timing for a space-based mid-infrared interferometer mission}
The idea for a space-based infrared interferometer for exoplanet science is not new. In fact, on both sides of the Atlantic, mission concepts were studied between the late 90s  and the mid 2000s (Darwin on ESA's side and TPF-I on NASA's side). {\bf While conceived as too risky and ahead of their time back then, the landscape for such a mission has completely changed -- as we will further detail below -- and the timing of ESA's ``Voyage 2050'' long-term plan could not be more ideal.} One of the most fundamental scientific results from the past years, with immediate relevance for the science discussed here, came from NASA's Kepler mission, namely that the statistical occurrence rate of small (terrestrial) planets  around solar-type and also M-type stars is extremely high, with -- on average -- more than one planet orbiting each star. This led to a number of new exoplanet missions to be proposed to both ESA and NASA, possibly culminating in a new exoplanet-driven flagship mission on the US side aiming at the direct detection of (small) exoplanets in reflected light (see section 3.2; see also White Paper by Snellen et al.). 
However, while the recently published ``Exoplanet Science Strategy" report from the US National Academies of Sciences\footnote{\url{https://www.nap.edu/catalog/25187/exoplanet-science-strategy}} puts a strong emphasis on future missions detecting planets in reflected light in a first step, mid-infrared interferometry is considered to be key in the long-run. One of the findings states: ``Technology development support in the next decade for future characterization concepts such as mid-infrared (MIR) interferometers [...] will be needed to enable strategic exoplanet missions beyond 2040." Even more important is the following statement: ``That said, the common (although often unspoken) belief is that such a nulling, near-infrared (NIR) interferometer would be a necessary follow-up to any reflected light direct imaging mission, as detecting the exoplanet in thermal emission is not only required to measure the temperature of the planet, but is also needed to measure its radius, and so (with an astrometric or radial velocity detection of [...] the mass of the planet) measure its density and thus determine if it is truly terrestrial.'' {\bf In combination with the results shown in Figure~\ref{fig:retrieval2}, underlining the unique characterization potential of thermal emission spectra, these statements re-emphasize not only the scientific importance, but also the timeliness for a space-based infrared nulling interferometer in the context of ESA's ``Voyage 2050'' long-term plan.} We note that it is not necessarily required that a reflected light mission has to be carried out first as shown by our yield analysis above. Finding and characterizing dozens of terrestrial exoplanets is feasible with a large MIR interferometer mission alone. It is, however, also clear that combining reflected light and thermal emission data for a given exoplanet expands the characterization potential significantly (see, e.g., section 1.2.1).

\section{Exoplanet science in the 2030s-2040s}
Since the first detection of a planet orbiting a main-sequence star other than our Sun \citep{mayor1995}, the field of exoplanet science has been growing at a breathtaking speed: to date we know more than 5000 exoplanets and exoplanet candidates\footnote{\url{https://exoplanets.nasa.gov}}. The overwhelming majority of these objects were detected via dedicated long-term surveys using indirect techniques (the RV or the transit technique) from both the ground (e.g., the HARPS survey or the California Planet Survey) and from space (e.g., NASA's Kepler mission). Thanks to the statistics derived from these surveys, we have a first quantitative understanding of the occurrence rate of different planet types as a function of their radius / mass, orbital period and also spectral type of the host star \citep[e.g.,][]{coughlin2016,mayor2011,bonfils2013}. In addition, we can put constraints on where the transition occurs between rock-dominated exoplanets and gas/atmosphere-dominated exoplanets \citep[e.g.,][]{rogers2015,wolfgang2016,chen2017}. The RV and transit techniques were also the techniques that revealed the first, rocky exoplanets orbiting within or close to the habitable zone of their (very) low-mass host stars located in the Solar neighborhood \citep{escude2016, gillon2017}. For some exoplanets, transit spectroscopy and/or secondary eclipse measurements (primarily done from space with the Hubble Space Telescope and the Spitzer Space Telescope) provide empirical constraints on their atmospheric composition \citep[e.g.,][]{seager2010,sing2016}. With a few exceptions \citep[e.g.,][]{kreidberg2014,dewit2018}, up to now these investigations targeted primarily so-called hot Jupiters, gas-giant planets on orbits with periods of a few days only. In the following we will summarize what developments from ground and space we can expect in the coming $\sim$20 years. The focus will be on developments with immediate relevance for the science proposed here, i.e., the direct detection and (atmospheric) characterization of terrestrial exoplanets using thermal emission spectra.  

\subsection{Expected developments on the ground}

{\bf RV:} Ongoing (large) programs with high-precision, high-resolution spectrographs (e.g., HARPS, HIRES) continue to search for and detect exoplanets over a range of masses and orbital periods. 
New spectrographs specifically designed to detect small, rocky planets either around nearby, red, low-mass stars (e.g., CARMENES, SPIRoU) or even aiming at reaching the detection threshold for an Earth-twin around a solar-type star of $\sim0.1$ m/s (e.g., ESPRESSO) are in operations now and have the potential to reveal new (small) exoplanets in the vicinity of the Sun (see also section 4.2). More spectrographs (e.g., NIRPS, HARPS3, EXPRES, NEID) with similar capabilities and science goals are currently in development and will support the search for low-mass, rocky planets in the solar neighborhood. Coupling a high-resolution spectrograph with an adaptive optics system on an 8\,m telescope may, in principle and under best conditions, allow for the detection of the nearest exoplanet, Proxima b, in reflected light \citep{lovis2017}. Similar science is expected for a few more objects with upcoming high-resolution spectrographs in the era of 30-40\,m Extremely Large Telescopes (ELTs) \citep[e.g.,][]{snellen2015,lopezmorales2019}. 


{\bf Transit:} The majority of ground-based exoplanet transit searches (e.g., WASP, KELT, MASCARA, HAT, TrES) are focusing on the detection of hot and warm gas giant planets. However, other projects  (e.g., NGTS, MEarth, Trappist, SPECULOOS) are designed to look for smaller and terrestrial exoplanets. The latter ones are only detectable around (very) low-mass stars, but can be located within their empirical habitable zone \citep{gillon2017}. Given the random orientation of planetary orbits in the plane of the sky, only a small minority of the existing exoplanets can be detected this way; most planets do \emph{not} transit. Consequently, for a given planet type, transiting planets have statistically a larger distance from the Sun compared to non-transiting planets rendering possible follow-up observations with, e.g.,  direct imaging techniques, more challenging.

{\bf High-contrast imaging:} All leading ground-based 8 m class observatories are equipped with high-contrast imaging instruments designed to directly detect massive gas giant planets ($>$2-3 M$_{\rm Jupiter}$ at large orbital separations ($>$20 AU) \citep[e.g.,][]{nielsen2019}. ESO and the Breakthrough Foundation just concluded a first experiment to directly detect thermal emission around 10 \textmu m from small planets around alpha Cen A with an upgraded MIR instrument at the VLT in Chile\footnote{\url{https://www.eso.org/public/news/eso1911/}}. However, even with 8 m telescopes, detecting thermal emission from a true Earth-analog around the nearest stars is prohibitively expensive in terms of observing time and only with the advent of the ELTs a few terrestrial exoplanets orbiting very nearby stars would come within reach if they existed \citep{quanz2015}. 


{\bf Microlensing:}  Large ground-based networks of dedicated telescopes continue to identify and monitor microlensing events and give access to a unique part of the exoplanet mass-separation parameter space as, in principle, they can constrain the occurrence rate of planets as a function of their mass (down Super-Earths) out to separations of around 10 AU \citep[e.g.,][]{cassan2012}. There is a strong detection bias towards finding planets around M-stars. The number of detections is still modest, compared to RV and transit searches, and hence the uncertainties are large, but in order to map out the exoplanet population -- in a statistical sense -- microlensing is indispensable. A drawback is that the majority of the events is located too far away from the Sun for any follow-up observations to be feasible. 

\subsection{Expected developments in space}
\begin{figure}[h!]
\centering
\includegraphics[width=0.99\linewidth]{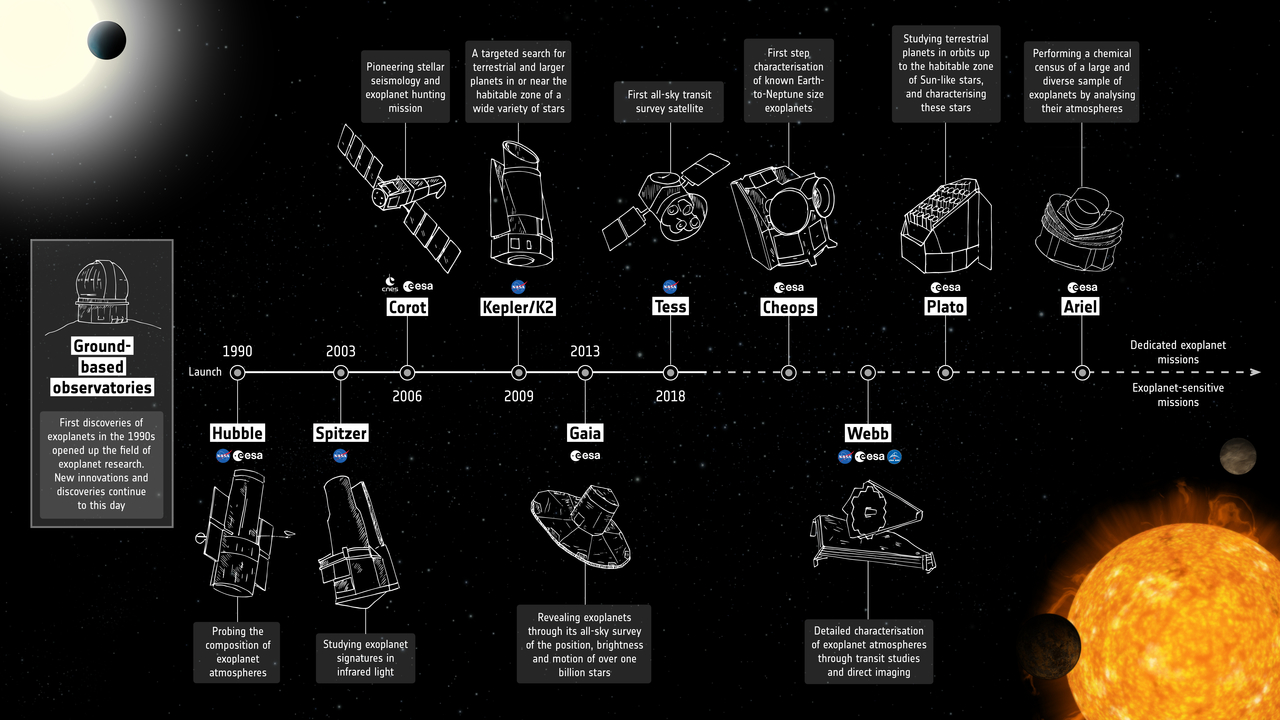}
    \caption{Adopted space missions related to exoplanet science from ESA. Image credit: ESA; \url{http://sci.esa.int/exoplanets/60657-the-future-of-exoplanet-research/} (accessed July 4, 2019)} 
    \label{fig:missionsESA}
\end{figure}

\begin{figure}[h!]
\centering
\includegraphics[width=0.88\linewidth]{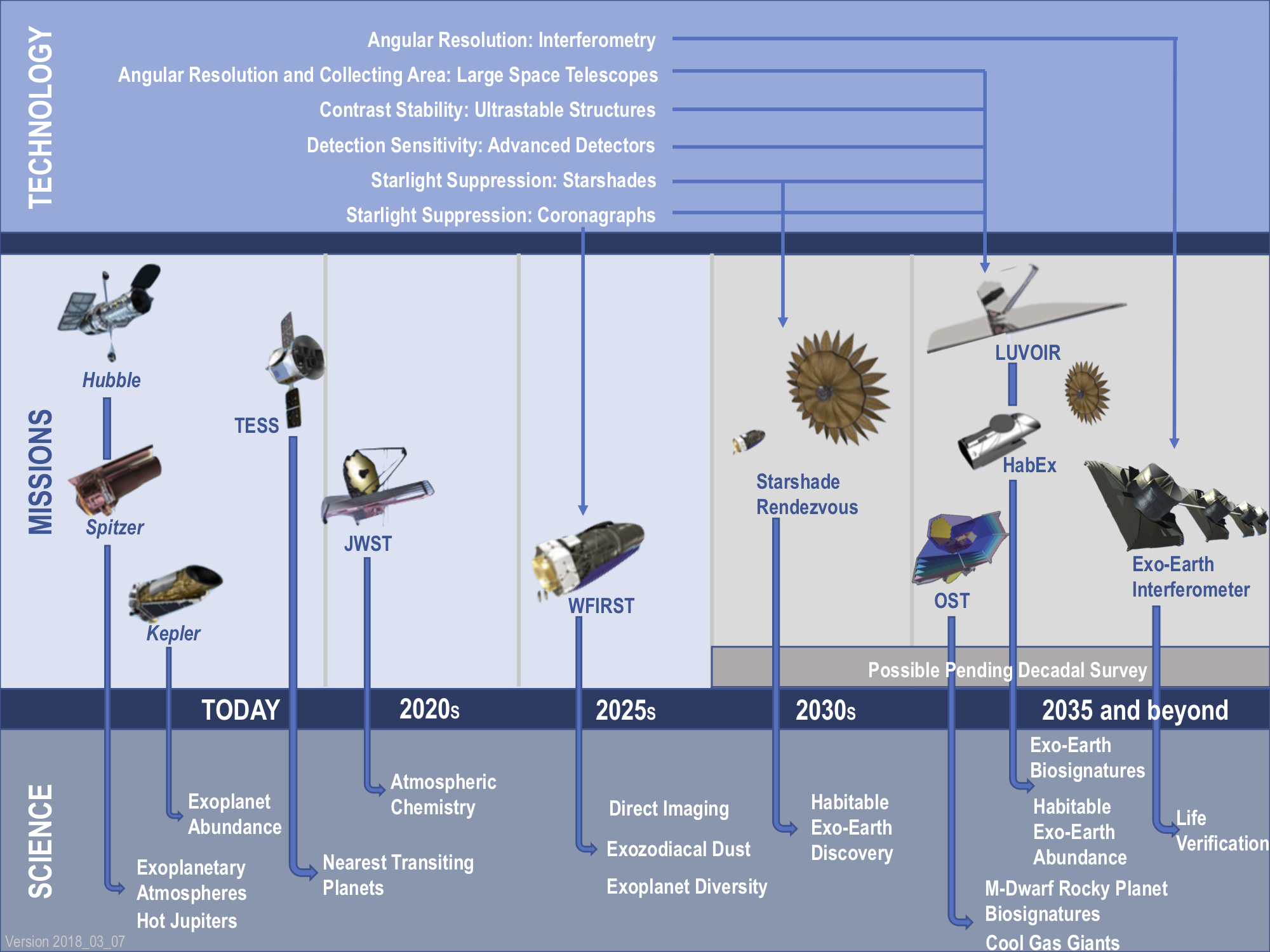}
    \caption{Space missions related to exoplanet science from NASA including  potential future missions that are part of the current Decadal Survey and also an MIR interferometer as discussed here. Image credit: NASA/JPL/Caltech; \url{https://exoplanets.nasa.gov/exep/technology/technology-overview/} (accessed July 4, 2019)} 
    \label{fig:missionsNASA}
\end{figure}

Both ESA and NASA are preparing to launch a suite of missions dedicated or related to exoplanet science in the coming 10-20 years that will join other already ongoing exoplanet missions (see Figures~\ref{fig:missionsESA} and \ref{fig:missionsNASA}). In addition to the dedicated exoplanet missions described below, ESA's Gaia mission will reveal thousands of  exoplanets based on the astrometric motion of their host stars providing a rich dataset for exoplanet population studies and targets for future imaging studies from ground and space \citep{sozzetti2018}. Whether Gaia can reveal Super-Earth exoplanets around some very nearby stars depends on the achievable astrometric accuracy at the end of the mission and remains to be seen. Not listed is ESA's EUCLID mission, which, while its primary mission is not exoplanet science, will deliver a large catalog of microlensing events, significantly extending the statistical power of this method.

{\bf ESA:} CHEOPS \citep[Characterizing Exoplanet Satellite;][]{cessa2017} is the first ESA S-class science mission with the goal of measuring the size of known transiting planets with high accuracy and searching for transit signals of well-selected exoplanets initially discovered with the RV technique. Launch is currently foreseen for end of 2019. In the mid 2020s, PLATO \citep[Planetary Transits and Oscillations of stars;][]{rauer2014} will follow as the third M-class mission in ESA's Cosmic Vision Program. Similar to Kepler, albeit targeting brighter stars with higher precision and longer time baseline, PLATO will uncover hundreds of new Earth-sized exoplanets and provide unprecedented constraints on the occurrence rate of terrestrial planets in the habitable zone of Solar-type stars. ARIEL \citep[Atmospheric Remote sensing Infrared Exoplanet Large survey mission;][]{tinetti2018}, another M-class exoplanet mission from ESA, will follow in 2028. ARIEL will provide transmission and secondary eclipse measurements for hundreds of (mostly transiting) exoplanets at visible and NIR wavelengths allowing investigations of the atmospheric composition of a large, well-defined and diverse sample of known exoplanets. The vast majority of ARIEL's targets will be warm and hot transiting gas giants and Neptunes. Some Super-Earths may also be within reach, but studying the atmospheres of temperate terrestrial exoplanets similar to Earth is beyond ARIEL's scope.

{\bf NASA:}  The currently operating missions Hubble, Spitzer, and the recently retired Kepler/K2 have revolutionized our understanding of exoplanet abundance and diversity through discovering and characterizing transiting systems. TESS launched successfully in 2018 and will bring another step forward in our understanding of exoplanet occurrence rates, especially around bright, nearby stars. Recently, NASA decided to extend the nominal 2-year mission lifetime by at least another 2 years significantly increasing TESS' discovery space\footnote{\url{https://science.nasa.gov/astrophysics/2019-senior-review-operating-missions}}. JWST (planned for a launch in 2021) will include the capability to perform infrared transit and eclipse spectroscopy of exoplanets as well as phase curve measurements \citep[e.g.,][]{greene2016}. The investigation of small planets orbiting close-to or within the habitable zone of their host stars will remain very challenging and time-demanding, though \citep[e.g.,][]{hedelt2013,kreidberg2016,morley2017}. WFIRST is planned for launch in the mid 2020s with a high-contrast coronagraph instrument (CGI). It will allow the direct detection of a few known giant exoplanets that were discovered by indirect techniques and perform an essential technology demonstration for future missions. A starshade could be launched to rendezvous with WFIRST, which would enable direct imaging of a few Earth-like exoplanets (pending recommendations of the 2020 Decadal Survey). Furthermore, the primary mission of WFIRST will, similar to EUCLID, deliver a wealth of microlensing events. If recommended, large missions like LUVOIR (Large UV/Optical/IR) Surveyor and HabEx (Habitable Exoplanet Observatory Mission) will be capable of directly imaging and spectrally characterizing up to a few tens of Earth-like exoplanets in reflected light \citep[][see also the White Paper by Snellen et al.]{kopparapu2018}. They may search for the spectral signature of gases like water vapor and oxygen \citep[see, Figure~\ref{fig:retrieval2};][]{feng2018}. The Origins Space Telescope (OST) plans to develop the capability to search for biosignature gases in the atmosphere of rocky exoplanets transiting M dwarfs. Any of these latter missions could be capable of discovering the first indications for signs of life on a nearby terrestrial exoplanet. However, they are all part of NASA's ongoing Decadal Survey and none of the missions is approved yet.  





\subsection{Implications for atmospheric studies of terrestrial exoplanets}
All of the currently adopted ground- or space-based projects and missions have exciting and challenging scientific objectives that will deliver important results in various areas of exoplanet research. However, none of them will enable the science proposed in this White Paper. JWST may be able to check for the existence of an atmosphere around a couple of nearby, terrestrial (transiting) exoplanets; a handful of terrestrial exoplanets may be also within reach of the ELTs for basic atmospheric characterization. As summarized on ESA's web-pages: ``With this suite of space telescopes launching within the next decade, we can expect to \emph{come closer} to finding Earth 2.0''\footnote{\url{http://sci.esa.int/exoplanets/60657-the-future-of-exoplanet-research/}}. However,  we will not have the means to find and characterize an Earth 2.0. Only a large and focused space mission offers the potential to do that and -- at the same time -- will allow us to statistically investigate the expected compositional diversity of terrestrial exoplanet atmospheres. 

\section{Challenges ahead}
Concepts for a space-based nulling interferometer already existed more than a decade ago, but technical challenges paired with uncertainties related to the scientific yield of such a mission -- the occurrence rate of small exoplanets was unknown -- led to the cancellation of the projects. Since then, progress has been made in  key areas, as we will detail in the following, but additional coordinated efforts will be needed to develop and space-qualify some components and technologies. Pushing the boundaries of what is technically possible always requires a substantial amount of investment, but only then scientific breakthrough results in (astro-)physics, such as the detection of gravitational waves \citep[e.g.,][]{grav_waves2016} or the first image of a black hole \citep{eventhorizon2019}, can be achieved. We picked these two examples on purpose because both of them relied on interferometric measurement techniques. We note that also ESA's LISA mission, bound to revolutionize our understanding of the universe using gravitational waves, will apply interferometric measurements between free-flying spacecraft. 

\subsection{Technology challenges and recent progress}
In recent years the field of high-precision ground-based interferometry has seen significant progress both in Europe and the United States. In particular Europe has gained a strong expertise in the field of fringe sensing, tracking, and stabilization with the operation of the Very Large Telescope Interferometer (VLTI). This maturity contributed to the first direct observation of an exoplanet with long-baseline optical interferometry, providing record-breaking precision on the astrometry and spectrum of any directly imaged planet to date \citep{Lacour2019}. In parallel, new data reduction and observing techniques have enabled unparalleled interferometric contrasts on US-based nulling interferometers \citep{hanot2011,defrere2016}. 

\begin{figure}[b!]
    \centering
\includegraphics[width=0.53\linewidth]{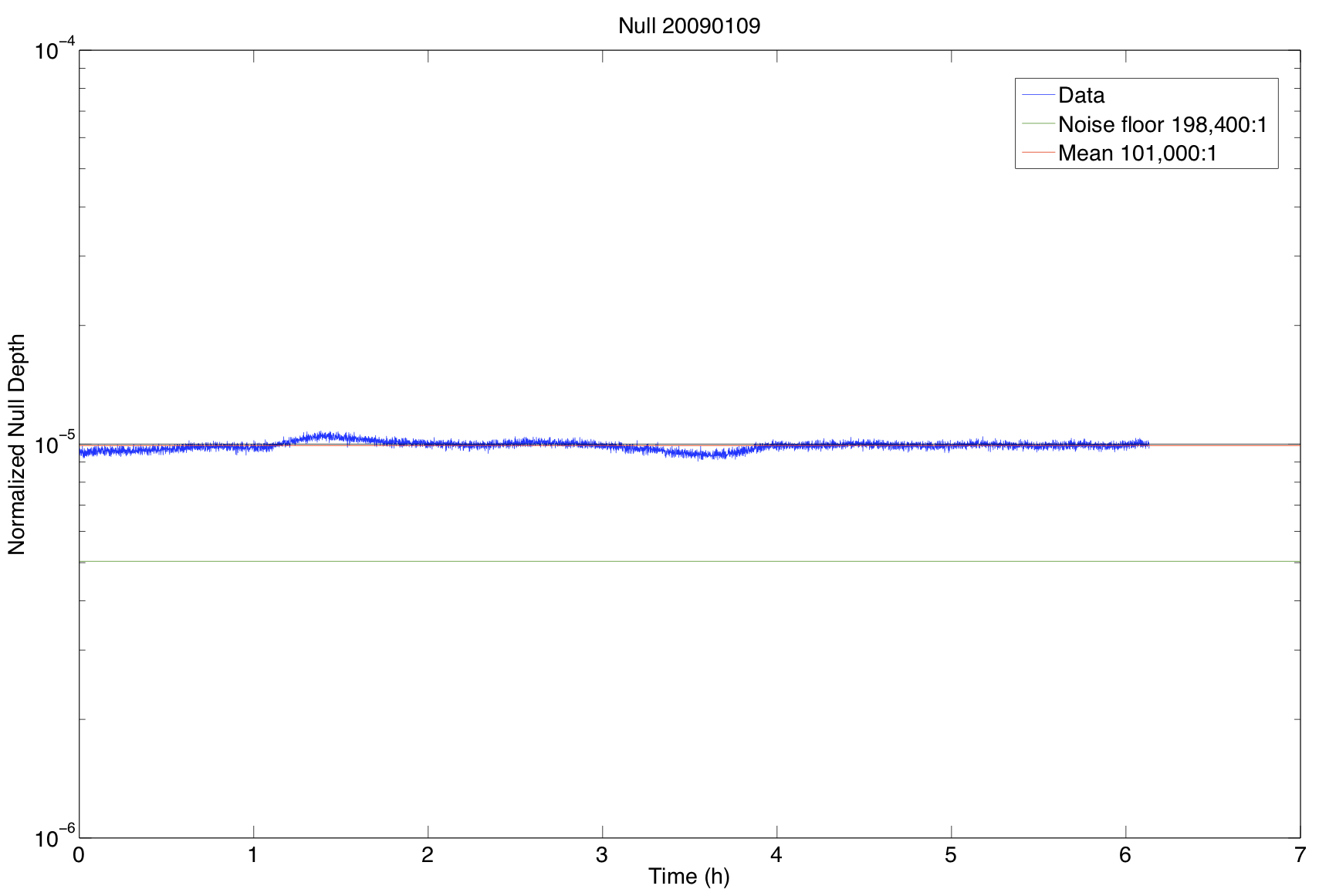}
\includegraphics[width=0.46\linewidth]{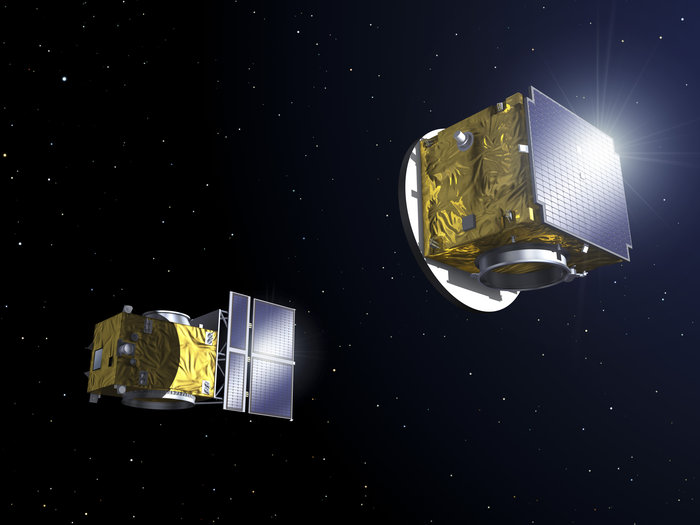}
    \caption{Left: 6-hour measurement of the null-depth achieved at 10~\textmu m wavelength with the ``Adaptive Nuller'' at JPL \citep{null_depth}. Right: Artist impression of ESA's PROBA 3 spacecraft, the first mission to demonstrate autonomous precision formation flying (Image credit: ESA - P. Carril, 2013).} 
    \label{fig:technology}
\end{figure}

In addition to these developments, most technologies required to fly a space-based nulling interferometer have reached a Technology Readiness Level (TRL) $\geq$5, which means that the components have been tested and validated in a relevant environment \citep[see, e.g., ][ for a recent review]{Defrere2018c}. In particular, key technologies that were considered immature in 2007, when most Darwin/TPF-I activities stopped, have now been demonstrated on test-benches (e.g., deep nulling beam combination) or will soon be demonstrated in space (e.g., formation flying). Significant efforts related to effective starlight suppression culminated in laboratory demonstrations mainly at the Jet Propulsion Laboratory (JPL) in the US. Work with the ``Adaptive Nuller'' at room temperature indicated that MIR nulls of 10$^{-5}$ are achievable with a bandwidth of 34\% and a mean wavelength of 10~\textmu m \citep[left panel in Figure~\ref{fig:technology};][]{null_depth}. The ``Planet detection testbed'', developed in parallel, {\bf demonstrated the main components of a high performance four beam nulling interferometer at a level matching that needed for a space mission} \citep{martin2012}. At 10~\textmu m, with 10\% bandwidth, it has achieved a null-depth of 8$\times$10$^{-6}$, and a total starlight suppression of 10$^{-8}$ after post-processing; the Earth-Sun contrast at 10~\textmu m is of the order 10$^{-7}$. 

Handling the high degree of autonomy necessary for free-flying or formation flying missions of close-by elements in space is a complex endeavor. This is an active field of research, with continuous progress in estimation and control algorithms, and in ways to internally calibrate the local time from the metrology to the scientific measurement systems \citep[e.g.,][and references therein]{linz2019}.
An additional key milestone for formation flying technology was the space-based demonstration by the PRISMA mission \citep{prisma}. 
PRISMA demonstrated a sub-cm positioning accuracy between two spacecraft, mainly limited by the metrology system (GPS and radio frequency ranging). The launch of ESA's PROBA-3 mission (right panel in Figure~\ref{fig:technology}), currently scheduled for late 2020, will mark the next step in formation flying. Its two satellites will maintain formation to millimeter and arc second precision at distances of 150 m or more autonomously, i.e., without relying on guidance from the ground. {\bf This separation is of the same order as the one needed for a space-based nulling interferometer (see above) and the formation flying precision exceeds even the requirements} \citep[cf.][]{defrere2018}. 

\subsection{Required technology development}
As mentioned above, the science described here will likely require an L-class mission and cost will be a key driver eventually. Hence, ideally, an ESA-supported technology development program should be set up that supports industrial partners and academic institutions in 
\begin{description}
    \item (1) identifying synergies with other planned or ongoing missions, leveraging the heritage of past (cryogenic / MIR) missions and investigating new approaches to limit major cost drivers during development and implementation of the new mission
    \item (2) further pushing the readiness and availability of certain required key technologies
\end{description}   

The general goal of cost reduction ensures that the total mission budget fits comfortably in the financial envelope of an L-class mission. Given that first mission concepts were already present 15-20 years ago and that significant further progress has been made ever since in various areas (see above), new concept studies would not have to start from scratch and could focus on identifying those areas where learnings from other missions or new developments and technologies would yield potential cost savings without jeopardizing the missions's scientific objectives. 

In addition, specific areas, where additional technology development would be required, are the following: To further push starlight suppression technology, the next step would be to reproduce the US experiments mentioned above, but under cryogenic conditions and with flux levels similar to those expected in space. The Laboratory for Astronomical Instrumentation at ETH Zurich\footnote{\url{https://quanz-group.ethz.ch/research/instrumentation.html}} is currently developing designs for possible experimental setups. This will likely include the successful validation of cryogenic spatial filters that can provide the necessary wave-front control performance from 3 to around 20~\textmu m and the implementation of a cryogenic deformable mirror \citep{zamkotsian2017,takahashi2017}. The use of newly available concepts of single-mode fibers as cryogenic spatial filters, including commercial solutions (e.g., classes of photonic crystal fibers, low-loss hollow core fibers), should be investigated in different ranges of the MIR spectrum because of their improved throughput across the spectral range of interest \citep[e.g.,][]{yu2012,sampaolo2015}. Furthermore,  integrated optics devices for the MIR wavelength regime could significantly reduce the complexity of the instrument if they reach the appropriate performance level. Recent developments seem promising \citep[e.g.,][]{tepper2017,labadie2018,gretzinger2019}, but more work is needed. 
We also expect that dedicated developments will be required in the field of MIR detectors, although the JWST legacy will be particularly useful in this context. However, also Mercury-Cadmium-Telluride (MCT) detectors seem to push towards longer wavelengths \citep{forrest2016} and it remains to be seen if this technology could reach out to at least $\sim$17~\textmu m to cover the important CO$_2$ band at 15~\textmu m (Figure~\ref{fig:molecules}).

\begin{figure}[t!]
    \centering
\includegraphics[width=0.6\linewidth]{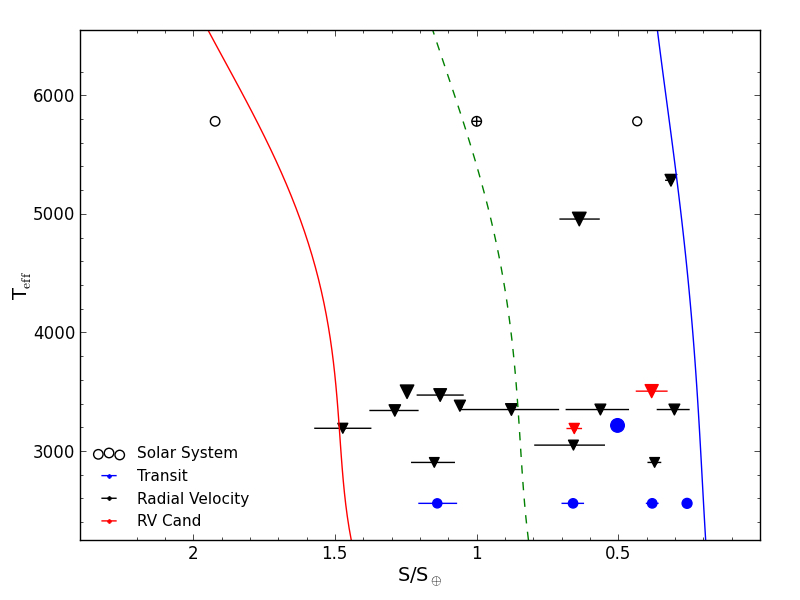}
    \caption{All known exoplanets and two additional yet unpublished candidates from the CARMENES project within 15 pc from the Sun and with masses  $<$10 M$_\oplus$ located within the empirical habitable zone of their host stars (boundaries indicated by the red and blue line; see section 1.2.2 for details). The x-axis shows the stellar insolation received by the exoplanets normalized to Earth's insolation. The y-axis shows the effective temperature of the host stars. Blue circles are planets detected via transit observations, black triangles are planets detected via RV measurements, and the unpublished candidates are shown in red. The locations of Venus, Earth and Mars are shown as black circles. 
    } 
    \label{fig:targets}
\end{figure}

\subsection{Astrophysical challenges}
A key factor impacting the time required to detect an ``Earth-like'' exoplanet at MIR wavelengths will be the level of thermal emission arising from small grains in exo-zodiacal dust belts around the target star. This emission would provide extra thermal background noise possibly leading to an increase in observing time for a given target and -- in the extreme case -- to an extension of the overall mission search phase. For the vast majority of stars within the immediate vicinity of the Sun ($<$ 15-20 pc) it is not known if they harbor such belts and -- if so -- what their level of thermal emission in the MIR range is. The NASA funded HOSTS survey, carried out with the Large Binocular Telescope Interferometer (LBTI), provides some new statistical results on the occurrence rate of exo-zodiacal dust, indicating that the median level for Sun-like stars amounts to $4.5^{+7.5}_{-1.5}$ times the level of the zodiacal light in the Solar System \citep[Ertel et al.\ (in prep);][]{ertel2018}. This work has shown that for a majority of systems the dust levels can expected to be low enough not to prevent the detection of terrestrial exoplanets through MIR interferometry. Ideally, however, a systematic observing program should be carried out to investigate all potential target stars of a future exoplanet imaging space mission including stars located in the Southern hemisphere that are inaccessible with the LBTI \citep[see, e.g., the Hi-5 project on the VLTI;][]{Defrere2018b}. In addition to the MIR flux levels of potential dust belts also their orientation / geometry would be good to know as this prior information can render the exoplanet search more efficient \citep[e.g.,][]{janson2010}.

Furthermore, to minimize the time devoted to an initial search phase for detecting a sufficiently large sample of terrestrial exoplanets for in-depth atmospheric characterization, current efforts to find such planets beforehand -- from the ground or from space -- need to be continued. The yield estimate shown in Figure~\ref{fig:yield} assumes that the planet occurrence rates detected by the Kepler mission are also applicable for exoplanets around stars in the Solar neighborhood. As shown in Figure~\ref{fig:targets}, close to 20 small exoplanets within 15 pc are already known that lie within the empirical habitable zone. Due to detection biases of current surveys such planets are preferably found around cooler, i.e., smaller and lower mass, M-stars. Not all of these planets will be rocky \citep[the transition between rocky and  envelope-dominated exoplanets, in terms of planet mass, is expected to occur roughly around (2.0$\pm$0.7)  M$_{\oplus}$;][]{chen2017} and some orbit too close around their host star so that even an MIR nulling interferometer, with the baselines and resulting spatial resolution as assumed above, could not directly detect them. However, these numbers show that planets orbiting in the empirical zone appear to be ubiquitous in the Solar neighborhood. We remind the reader that there are of order 900 GKM main-sequence stars within 15 pc from the Sun (see, Figure~\ref{fig:stars}), many of which have not yet been searched for (terrestrial) exoplanets. Recent estimates for $\eta_\oplus$, i.e., fraction of stars with terrestrial planets located within the habitable zone, amount to roughly 0.2-0.3 for solar type stars and possibly even higher values for M-stars \citep[e.g.,][]{kaltenegger2017}. New, dedicated RV projects specifically targeting exoplanets close-to and in the habitable zone around nearby G- and K-type stars (such as, e.g., EXPRES\footnote{\url{http://exoplanets.astro.yale.edu/instrumentation/expres.php}} or HARPS3\footnote{\url{http://www.terrahunting.org/index.html}}) would be perfectly complementary to ongoing projects focusing on low-mass stars. Combined they will not only provide targets for a future exoplanet imaging space mission, but also crucial constraints on the planets' masses.

\section{Conclusions}
Exoplanet science is one of the most thriving fields of modern astrophysics. A major goal is the atmospheric characterization of dozens of small, terrestrial exoplanets in order to search for signatures in their atmospheres that indicate biological activity, assess their ability to provide conditions for life as we know it, and investigate their expected atmospheric diversity. None of the currently adopted projects or missions, from ground or in space, can address these goals. In this White Paper we have argued that a large space-based mission designed to {\bf detect and investigate thermal emission spectra of terrestrial exoplanets in the MIR wavelength range provides  unique scientific potential to address these goals and surpasses the capabilities of other approaches}. While NASA might be focusing on large missions that aim to detect terrestrial planets in reflected light, {\bf ESA has the opportunity to take leadership and spearhead the development of a large MIR exoplanet mission within the scope of the  ``Voyage 2050'' long-term plan establishing Europe at the forefront of exoplanet science for decades to come.} Given the ambitious science goals of such a mission, additional international partners might be interested in participating and contributing to a roadmap that, in the long run, leads to a successful implementation. A new, dedicated development program funded by ESA to help reduce development and implementation cost and further push some of the required key technologies would be a first important step in this direction. 

Ultimately, a large MIR exoplanet imaging mission will be needed to help answer one of mankind's most fundamental questions: ``How unique is our Earth?''\newpage
\bibliographystyle{apj} 

\begin{thebibliography}{}
\expandafter\ifx\csname natexlab\endcsname\relax\def\natexlab#1{#1}\fi

\bibitem[{Abbott {et~al.}(2016)Abbott, Abbott, Abbott, Abernathy, Acernese,
  Ackley, Adams, Adams, Addesso, Adhikari, Adya, Affeldt, Agathos, Agatsuma,
  Aggarwal, Aguiar, Aiello, Ain, Ajith, Allen, Allocca, Altin, Anderson,
  Anderson, Arai, Arain, Araya, Arceneaux, Areeda, Arnaud, Arun, Ascenzi,
  Ashton, Ast, Aston, Astone, Aufmuth, Aulbert, Babak, Bacon, Bader, Baker,
  Baldaccini, Ballardin, Ballmer, Barayoga, Barclay, Barish, Barker, Barone,
  Barr, Barsotti, Barsuglia, Barta, Bartlett, Barton, Bartos, Bassiri, Basti,
  Batch, Baune, Bavigadda, Bazzan, Behnke, Bejger, Belczynski, Bell, Bell,
  Berger, Bergman, Bergmann, Berry, Bersanetti, Bertolini, Betzwieser, Bhagwat,
  Bhandare, Bilenko, Billingsley, Birch, Birney, Birnholtz, Biscans, Bisht,
  Bitossi, Biwer, Bizouard, Blackburn, Blair, Blair, Blair, Bloemen, Bock,
  Bodiya, Boer, Bogaert, Bogan, Bohe, Bojtos, Bond, Bondu, Bonnand, Boom, Bork,
  Boschi, Bose, Bouffanais, Bozzi, Bradaschia, Brady, Braginsky, Branchesi,
  Brau, Briant, Brillet, Brinkmann, Brisson, Brockill, Brooks, Brown, Brown,
  Brown, Buchanan, Buikema, Bulik, Bulten, Buonanno, Buskulic, Buy, Byer,
  Cabero, Cadonati, Cagnoli, Cahillane, Bustillo, Callister, Calloni, Camp,
  Cannon, Cao, Capano, Capocasa, Carbognani, Caride, Diaz, Casentini, Caudill,
  Cavagli\`a, Cavalier, Cavalieri, Cella, Cepeda, Baiardi, Cerretani, Cesarini,
  Chakraborty, Chalermsongsak, Chamberlin, Chan, Chao, Charlton,
  Chassande-Mottin, Chen, Chen, Cheng, Chincarini, Chiummo, Cho, Cho, Chow,
  Christensen, Chu, Chua, Chung, Ciani, Clara, Clark, Cleva, Coccia, Cohadon,
  Colla, Collette, Cominsky, Constancio, Conte, Conti, Cook, Corbitt, Cornish,
  Corsi, Cortese, Costa, Coughlin, Coughlin, Coulon, Countryman, Couvares,
  Cowan, Coward, Cowart, Coyne, Coyne, Craig, Creighton, Creighton, Cripe,
  Crowder, Cruise, Cumming, Cunningham, Cuoco, Canton, Danilishin, D'Antonio,
  Danzmann, Darman, Da~Silva~Costa, Dattilo, Dave, Daveloza, Davier, Davies,
  Daw, Day, De, DeBra, Debreczeni, Degallaix, De~Laurentis, Del\'eglise,
  Del~Pozzo, Denker, Dent, Dereli, Dergachev, DeRosa, De~Rosa, DeSalvo,
  Dhurandhar, D\'{\i}az, Di~Fiore, Di~Giovanni, Di~Lieto, Di~Pace, Di~Palma,
  Di~Virgilio, Dojcinoski, Dolique, Donovan, Dooley, Doravari, Douglas, Downes,
  Drago, Drever, Driggers, Du, Ducrot, Dwyer, Edo, Edwards, Effler, Eggenstein,
  Ehrens, Eichholz, Eikenberry, Engels, Essick, Etzel, Evans, Evans, Everett,
  Factourovich, Fafone, Fair, Fairhurst, Fan, Fang, Farinon, Farr, Farr,
  Favata, Fays, Fehrmann, Fejer, Feldbaum, Ferrante, Ferreira, Ferrini,
  Fidecaro, Finn, Fiori, Fiorucci, Fisher, Flaminio, Fletcher, Fong, Fournier,
  Franco, Frasca, Frasconi, Frede, Frei, Freise, Frey, Frey, Fricke, Fritschel,
  Frolov, Fulda, Fyffe, Gabbard, Gair, Gammaitoni, Gaonkar, Garufi, Gatto,
  Gaur, Gehrels, Gemme, Gendre, Genin, Gennai, George, Gergely, Germain, Ghosh,
  Ghosh, Ghosh, Giaime, Giardina, Giazotto, Gill, Glaefke, Gleason, Goetz,
  Goetz, Gondan, Gonz\'alez, Castro, Gopakumar, Gordon, Gorodetsky, Gossan,
  Gosselin, Gouaty, Graef, Graff, Granata, Grant, Gras, Gray, Greco, Green,
  Greenhalgh, Groot, Grote, Grunewald, Guidi, Guo, Gupta, Gupta, Gushwa,
  Gustafson, Gustafson, Hacker, Hall, Hall, Hammond, Haney, Hanke, Hanks,
  Hanna, Hannam, Hanson, Hardwick, Harms, Harry, Harry, Hart, Hartman, Haster,
  Haughian, Healy, Heefner, Heidmann, Heintze, Heinzel, Heitmann, Hello,
  Hemming, Hendry, Heng, Hennig, Heptonstall, Heurs, Hild, Hoak, Hodge, Hofman,
  Hollitt, Holt, Holz, Hopkins, Hosken, Hough, Houston, Howell, Hu, Huang,
  Huerta, Huet, Hughey, Husa, Huttner, Huynh-Dinh, Idrisy, Indik, Ingram, Inta,
  Isa, Isac, Isi, Islas, Isogai, Iyer, Izumi, Jacobson, Jacqmin, Jang, Jani,
  Jaranowski, Jawahar, Jim\'enez-Forteza, Johnson, Johnson-McDaniel, Jones,
  Jones, Jonker, Ju, Haris, Kalaghatgi, Kalogera, Kandhasamy, Kang, Kanner,
  Karki, Kasprzack, Katsavounidis, Katzman, Kaufer, Kaur, Kawabe, Kawazoe,
  K\'ef\'elian, Kehl, Keitel, Kelley, Kells, Kennedy, Keppel, Key,
  Khalaidovski, Khalili, Khan, Khan, Khan, Khazanov, Kijbunchoo, Kim, Kim, Kim,
  Kim, Kim, Kim, King, King, Kinzel, Kissel, Kleybolte, Klimenko, Koehlenbeck,
  Kokeyama, Koley, Kondrashov, Kontos, Koranda, Korobko, Korth, Kowalska,
  Kozak, Kringel, Krishnan, Kr\'olak, Krueger, Kuehn, Kumar, Kumar, Kuo,
  Kutynia, Kwee, Lackey, Landry, Lange, Lantz, Lasky, Lazzarini, Lazzaro,
  Leaci, Leavey, Lebigot, Lee, Lee, Lee, Lee, Lenon, Leonardi, Leong, Leroy,
  Letendre, Levin, Levine, Li, Libson, Littenberg, Lockerbie, Logue, Lombardi,
  London, Lord, Lorenzini, Loriette, Lormand, Losurdo, Lough, Lousto, Lovelace,
  L\"uck, Lundgren, Luo, Lynch, Ma, MacDonald, Machenschalk, MacInnis, Macleod,
  Maga\~na Sandoval, Magee, Mageswaran, Majorana, Maksimovic, Malvezzi, Man,
  Mandel, Mandic, Mangano, Mansell, Manske, Mantovani, Marchesoni, Marion,
  M\'arka, M\'arka, Markosyan, Maros, Martelli, Martellini, Martin, Martin,
  Martynov, Marx, Mason, Masserot, Massinger, Masso-Reid, Matichard, Matone,
  Mavalvala, Mazumder, Mazzolo, McCarthy, McClelland, McCormick, McGuire,
  McIntyre, McIver, McManus, McWilliams, Meacher, Meadors, Meidam, Melatos,
  Mendell, Mendoza-Gandara, Mercer, Merilh, Merzougui, Meshkov, Messenger,
  Messick, Meyers, Mezzani, Miao, Michel, Middleton, Mikhailov, Milano, Miller,
  Millhouse, Minenkov, Ming, Mirshekari, Mishra, Mitra, Mitrofanov,
  Mitselmakher, Mittleman, Moggi, Mohan, Mohapatra, Montani, Moore, Moore,
  Moraru, Moreno, Morriss, Mossavi, Mours, Mow-Lowry, Mueller, Mueller, Muir,
  Mukherjee, Mukherjee, Mukherjee, Mukund, Mullavey, Munch, Murphy, Murray,
  Mytidis, Nardecchia, Naticchioni, Nayak, Necula, Nedkova, Nelemans, Neri,
  Neunzert, Newton, Nguyen, Nielsen, Nissanke, Nitz, Nocera, Nolting,
  Normandin, Nuttall, Oberling, Ochsner, O'Dell, Oelker, Ogin, Oh, Oh, Ohme,
  Oliver, Oppermann, Oram, O'Reilly, O'Shaughnessy, Ott, Ottaway, Ottens,
  Overmier, Owen, Pai, Pai, Palamos, Palashov, Palomba, Pal-Singh, Pan, Pan,
  Pankow, Pannarale, Pant, Paoletti, Paoli, Papa, Paris, Parker, Pascucci,
  Pasqualetti, Passaquieti, Passuello, Patricelli, Patrick, Pearlstone,
  Pedraza, Pedurand, Pekowsky, Pele, Penn, Perreca, Pfeiffer, Phelps, Piccinni,
  Pichot, Pickenpack, Piergiovanni, Pierro, Pillant, Pinard, Pinto, Pitkin,
  Poeld, Poggiani, Popolizio, Post, Powell, Prasad, Predoi, Premachandra,
  Prestegard, Price, Prijatelj, Principe, Privitera, Prix, Prodi, Prokhorov,
  Puncken, Punturo, Puppo, P\"urrer, Qi, Qin, Quetschke, Quintero,
  Quitzow-James, Raab, Rabeling, Radkins, Raffai, Raja, Rakhmanov, Ramet,
  Rapagnani, Raymond, Razzano, Re, Read, Reed, Regimbau, Rei, Reid, Reitze,
  Rew, Reyes, Ricci, Riles, Robertson, Robie, Robinet, Rocchi, Rolland,
  Rollins, Roma, Romano, Romano, Romanov, Romie, Rosi\ifmmode~\acute{n}\else
  \'{n}\fi{}ska, Rowan, R\"udiger, Ruggi, Ryan, Sachdev, Sadecki, Sadeghian,
  Salconi, Saleem, Salemi, Samajdar, Sammut, Sampson, Sanchez, Sandberg,
  Sandeen, Sanders, Sanders, Sassolas, Sathyaprakash, Saulson, Sauter, Savage,
  Sawadsky, Schale, Schilling, Schmidt, Schmidt, Schnabel, Schofield,
  Sch\"onbeck, Schreiber, Schuette, Schutz, Scott, Scott, Sellers, Sengupta,
  Sentenac, Sequino, Sergeev, Serna, Setyawati, Sevigny, Shaddock, Shaffer,
  Shah, Shahriar, Shaltev, Shao, Shapiro, Shawhan, Sheperd, Shoemaker,
  Shoemaker, Siellez, Siemens, Sigg, Silva, Simakov, Singer, Singer, Singh,
  Singh, Singhal, Sintes, Slagmolen, Smith, Smith, Smith, Smith, Son, Sorazu,
  Sorrentino, Souradeep, Srivastava, Staley, Steinke, Steinlechner,
  Steinlechner, Steinmeyer, Stephens, Stevenson, Stone, Strain, Straniero,
  Stratta, Strauss, Strigin, Sturani, Stuver, Summerscales, Sun, Sutton,
  Swinkels, Szczepa\ifmmode~\acute{n}\else \'{n}\fi{}czyk, Tacca, Talukder,
  Tanner, T\'apai, Tarabrin, Taracchini, Taylor, Theeg, Thirugnanasambandam,
  Thomas, Thomas, Thomas, Thorne, Thorne, Thrane, Tiwari, Tiwari, Tokmakov,
  Tomlinson, Tonelli, Torres, Torrie, T\"oyr\"a, Travasso, Traylor, Trifir\`o,
  Tringali, Trozzo, Tse, Turconi, Tuyenbayev, Ugolini, Unnikrishnan, Urban,
  Usman, Vahlbruch, Vajente, Valdes, Vallisneri, van Bakel, van Beuzekom,
  van~den Brand, Van Den~Broeck, Vander-Hyde, van~der Schaaf, van Heijningen,
  van Veggel, Vardaro, Vass, Vas\'uth, Vaulin, Vecchio, Vedovato, Veitch,
  Veitch, Venkateswara, Verkindt, Vetrano, Vicer\'e, Vinciguerra, Vine, Vinet,
  Vitale, Vo, Vocca, Vorvick, Voss, Vousden, Vyatchanin, Wade, Wade, Wade,
  Waldman, Walker, Wallace, Walsh, Wang, Wang, Wang, Wang, Wang, Ward, Ward,
  Warner, Was, Weaver, Wei, Weinert, Weinstein, Weiss, Welborn, Wen,
  We\ss{}els, Westphal, Wette, Whelan, Whitcomb, White, Whiting, Wiesner,
  Wilkinson, Willems, Williams, Williams, Williamson, Willis, Willke, Wimmer,
  Winkelmann, Winkler, Wipf, Wiseman, Wittel, Woan, Worden, Wright, Wu, Yablon,
  Yakushin, Yam, Yamamoto, Yancey, Yap, Yu, Yvert, Zadro\ifmmode~\dot{z}\else
  \.{z}\fi{}ny, Zangrando, Zanolin, Zendri, Zevin, Zhang, Zhang, Zhang, Zhang,
  Zhao, Zhou, Zhou, Zhu, Zucker, Zuraw, \& Zweizig}]{grav_waves2016}
Abbott, B.~P., Abbott, R., Abbott, T.~D., {et~al.} 2016, Phys. Rev. Lett., 116,
  061102

\bibitem[{{Absil} {et~al.}(2006){Absil}, {den Hartog}, {Gondoin}, {Fabry},
  {Wilhelm}, {Gitton}, \& {Puech}}]{absil2006}
{Absil}, O., {den Hartog}, R., {Gondoin}, P., {et~al.} 2006, \aap, 448, 787

\bibitem[{Anglada-Escud{\'e} {et~al.}(2016)Anglada-Escud{\'e}, Amado, Barnes,
  Berdi{\~n}as, Butler, Coleman, de~la Cueva, Dreizler, Endl, Giesers, Jeffers,
  Jenkins, Jones, Kiraga, K{\"u}rster, L{\'o}pez-Gonz{\'a}lez, Marvin, Morales,
  Morin, Nelson, Ortiz, Ofir, Paardekooper, Reiners, Rodr{\'\i}guez,
  Rodriguez-L{\'o}pez, Sarmiento, Strachan, Tsapras, Tuomi, \&
  Zechmeister}]{escude2016}
Anglada-Escud{\'e}, G., Amado, P.~J., Barnes, J., {et~al.} 2016, Nature, 536, 1

\bibitem[{Bodin {et~al.}(2012)Bodin, Noteborn, Larsson, Karlsson, D{'}Amico,
  Ardaens, Delpech, \& Berges}]{prisma}
Bodin, P., Noteborn, R., Larsson, R., {et~al.} 2012, in 35th Annual AAS
  Guidance and Control Conference

\bibitem[{{Bonfils} {et~al.}(2013){Bonfils}, {Delfosse}, {Udry}, {Forveille},
  {Mayor}, {Perrier}, {Bouchy}, {Gillon}, {Lovis}, \& {Pepe}}]{bonfils2013}
{Bonfils}, X., {Delfosse}, X., {Udry}, S., {et~al.} 2013, \aap, 549, A109

\bibitem[{{Burke} {et~al.}(2015){Burke}, {Christiansen}, {Mullally}, {Seader},
  {Huber}, {Rowe}, {Coughlin}, {Thompson}, {Catanzarite}, {Clarke}, {Morton},
  {Caldwell}, {Bryson}, {Haas}, {Batalha}, {Jenkins}, {Tenenbaum}, {Twicken},
  {Li}, {Quintana}, {Barclay}, {Henze}, {Borucki}, {Howell}, \&
  {Still}}]{burke2015}
{Burke}, C.~J., {Christiansen}, J.~L., {Mullally}, F., {et~al.} 2015, \apj,
  809, 8

\bibitem[{{Cassan} {et~al.}(2012){Cassan}, {Kubas}, {Beaulieu}, {Dominik},
  {Horne}, {Greenhill}, {Wambsganss}, {Menzies}, {Williams}, {J{\o}rgensen},
  {Udalski}, {Bennett}, {Albrow}, {Batista}, {Brillant}, {Caldwell}, {Cole},
  {Coutures}, {Cook}, {Dieters}, {Dominis Prester}, {Donatowicz}, {Fouqu{\'e}},
  {Hill}, {Kains}, {Kane}, {Marquette}, {Martin}, {Pollard}, {Sahu}, {Vinter},
  {Warren}, {Watson}, {Zub}, {Sumi}, {Szyma{\'n}ski}, {Kubiak}, {Poleski},
  {Soszynski}, {Ulaczyk}, {Pietrzy{\'n}ski}, \& {Wyrzykowski}}]{cassan2012}
{Cassan}, A., {Kubas}, D., {Beaulieu}, J.~P., {et~al.} 2012, \nat, 481, 167

\bibitem[{{Catling} {et~al.}(2018){Catling}, {Krissansen-Totton}, {Kiang},
  {Crisp}, {Robinson}, {DasSarma}, {Rushby}, {Del Genio}, {Bains}, \&
  {Domagal-Goldman}}]{catling2018}
{Catling}, D.~C., {Krissansen-Totton}, J., {Kiang}, N.~Y., {et~al.} 2018,
  Astrobiology, 18, 709

\bibitem[{{Cessa} {et~al.}(2017){Cessa}, {Beck}, {Benz}, {Broeg}, {Ehrenreich},
  {Fortier}, {Peter}, {Magrin}, {Pagano}, \& {Plesseria}}]{cessa2017}
{Cessa}, V., {Beck}, T., {Benz}, W., {et~al.} 2017, in Society of Photo-Optical
  Instrumentation Engineers (SPIE) Conference Series, Vol. 10563, SPIE, 105631L

\bibitem[{Chen \& Kipping(2017)}]{chen2017}
Chen, J., \& Kipping, D. 2017, The Astrophysical Journal, 834, 17

\bibitem[{Cockell {et~al.}(2009)Cockell, Herbst, L{\'e}ger, Absil, Beichman,
  Benz, Brack, Chazelas, Chelli, Cottin, Coud{\'e}~du Foresto, Danchi,
  Defr{\`e}re, den Herder, Eiroa, Fridlund, Henning, Johnston, Kaltenegger,
  Labadie, Lammer, Launhardt, Lawson, Lay, Liseau, Martin, Mawet, Mourard,
  Moutou, Mugnier, Paresce, Quirrenbach, Rabbia, Rottgering, Rouan, Santos,
  Selsis, Serabyn, Westall, White, Ollivier, \& Bord{\'e}}]{cockell2009}
Cockell, C.~S., Herbst, T., L{\'e}ger, A., {et~al.} 2009, Experimental
  Astronomy, 23, 435

\bibitem[{Coughlin {et~al.}(2016)Coughlin, Mullally, Thompson, Rowe, Burke,
  Latham, Batalha, Ofir, Quarles, Henze, Wolfgang, Caldwell, Bryson, Shporer,
  Catanzarite, Akeson, Barclay, Borucki, Boyajian, Campbell, Christiansen,
  Girouard, Haas, Howell, Huber, Jenkins, Li, Patil-Sabale, Quintana, Ramirez,
  Seader, Smith, Tenenbaum, Twicken, \& Zamudio}]{coughlin2016}
Coughlin, J.~L., Mullally, F., Thompson, S.~E., {et~al.} 2016, The
  Astrophysical Journal Supplement Series, 224, 12

\bibitem[{Crossfield(2013)}]{crossfield2013}
Crossfield, I. J.~M. 2013, Astronomy {\&} Astrophysics, 551, A99

\bibitem[{de~Wit {et~al.}(2018)de~Wit, Wakeford, Lewis, Delrez, Gillon, Selsis,
  Leconte, Demory, Bolmont, Bourrier, Burgasser, Grimm, Jehin, Lederer, Owen,
  Stamenkovi{\'c}, \& Triaud}]{dewit2018}
de~Wit, J., Wakeford, H.~R., Lewis, N.~K., {et~al.} 2018, Nature Astronomy, 2,
  214

\bibitem[{{Defr{\`e}re} {et~al.}(2018){Defr{\`e}re}, {Absil}, \&
  {Beichman}}]{Defrere2018c}
{Defr{\`e}re}, D., {Absil}, O., \& {Beichman}, C.~A. 2018, {Interferometric
  Space Missions for Exoplanet Science: Legacy of Darwin/TPF}, 82

\bibitem[{{Defr{\`e}re} {et~al.}(2016){Defr{\`e}re}, {Hinz}, {Mennesson},
  {Hoffmann}, {Millan-Gabet}, {Skemer}, {Bailey}, {Danchi}, {Downey}, {Durney},
  {Grenz}, {Hill}, {McMahon}, {Montoya}, {Spalding}, {Vaz}, {Absil}, {Arbo},
  {Bailey}, {Brusa}, {Bryden}, {Esposito}, {Gaspar}, {Haniff}, {Kennedy},
  {Leisenring}, {Marion}, {Nowak}, {Pinna}, {Powell}, {Puglisi}, {Rieke},
  {Roberge}, {Serabyn}, {Sosa}, {Stapeldfeldt}, {Su}, {Weinberger}, \&
  {Wyatt}}]{defrere2016}
{Defr{\`e}re}, D., {Hinz}, P.~M., {Mennesson}, B., {et~al.} 2016, \apj, 824, 66

\bibitem[{Defr{\`e}re {et~al.}(2018)Defr{\`e}re, L{\'e}ger, Absil, Beichman,
  Biller, Danchi, Ergenzinger, Eiroa, Ertel, Fridlund, noz, Gillon, Glasse11,
  Godolt, Grenfell, Kraus, Labadie, Lacour, Liseau, Martin, Mennesson, Micela,
  Minardi, Quanz, Rauer, Rinehart, Santos, Selsis, Surdej, Tian, Villaver,
  Wheatley, \& Wyatt}]{defrere2018}
Defr{\`e}re, D., L{\'e}ger, A., Absil, O., {et~al.} 2018, Experimental
  Astronomy, 7, 1

\bibitem[{{Defr{\`e}re} {et~al.}(2018){Defr{\`e}re}, {Absil}, {Berger},
  {Boulet}, {Danchi}, {Ertel}, {Gallenne}, {H{\'e}nault}, {Hinz}, {Huby},
  {Ireland}, {Kraus}, {Labadie}, {Le Bouquin}, {Martin}, {Matter},
  {M{\'e}rand}, {Mennesson}, {Minardi}, {Monnier}, {Norris}, {de Xivry},
  {Pedretti}, {Pott}, {Reggiani}, {Serabyn}, {Surdej}, {Tristram}, \&
  {Woillez}}]{Defrere2018b}
{Defr{\`e}re}, D., {Absil}, O., {Berger}, J.-P., {et~al.} 2018, Experimental
  Astronomy, 46, 475

\bibitem[{{Des Marais} {et~al.}(2002){Des Marais}, {Harwit}, {Jucks},
  {Kasting}, {Lin}, {Lunine}, {Schneider}, {Seager}, {Traub}, \&
  {Woolf}}]{desmarais2002}
{Des Marais}, D.~J., {Harwit}, M.~O., {Jucks}, K.~W., {et~al.} 2002,
  Astrobiology, 2, 153

\bibitem[{Dressing \& Charbonneau(2015)}]{dressing2015}
Dressing, C.~D., \& Charbonneau, D. 2015, The Astrophysical Journal, 807, 45

\bibitem[{{Ertel} {et~al.}(2018){Ertel}, {Defr{\`e}re}, {Hinz}, {Mennesson},
  {Kennedy}, {Danchi}, {Gelino}, {Hill}, {Hoffmann}, {Rieke}, {Shannon},
  {Spalding}, {Stone}, {Vaz}, {Weinberger}, {Willems}, {Absil}, {Arbo},
  {Bailey}, {Beichman}, {Bryden}, {Downey}, {Durney}, {Esposito}, {Gaspar},
  {Grenz}, {Haniff}, {Leisenring}, {Marion}, {McMahon}, {Millan-Gabet},
  {Montoya}, {Morzinski}, {Pinna}, {Power}, {Puglisi}, {Roberge}, {Serabyn},
  {Skemer}, {Stapelfeldt}, {Su}, {Vaitheeswaran}, \& {Wyatt}}]{ertel2018}
{Ertel}, S., {Defr{\`e}re}, D., {Hinz}, P., {et~al.} 2018, \aj, 155, 194

\bibitem[{{Event Horizon Telescope Collaboration} {et~al.}(2019){Event Horizon
  Telescope Collaboration}, {Akiyama}, {Alberdi}, {Alef}, {Asada}, {Azulay},
  {Baczko}, {Ball}, {Balokovi{\'c}}, {Barrett}, \& et~al.}]{eventhorizon2019}
{Event Horizon Telescope Collaboration}, {Akiyama}, K., {Alberdi}, A., {et~al.}
  2019, \apjl, 875, L1

\bibitem[{Feng {et~al.}(2018)Feng, Robinson, Fortney, Lupu, Marley, Lewis,
  Macintosh, \& Line}]{feng2018}
Feng, Y.~K., Robinson, T.~D., Fortney, J.~J., {et~al.} 2018, The Astronomical
  Journal, 155, 200

\bibitem[{{Forrest} {et~al.}(2016){Forrest}, {McMurtry}, {Dorn}, {Pipher}, \&
  {Cabrera}}]{forrest2016}
{Forrest}, W.~J., {McMurtry}, C.~W., {Dorn}, M., {Pipher}, J., \& {Cabrera},
  M.~S. 2016, in AAS/Division for Planetary Sciences Meeting Abstracts \#48,
  AAS/Division for Planetary Sciences Meeting Abstracts, 123.59

\bibitem[{{Gebauer} {et~al.}(2017){Gebauer}, {Grenfell}, {Stock}, {Lehmann},
  {Godolt}, {von Paris}, \& {Rauer}}]{gebauer2017}
{Gebauer}, S., {Grenfell}, J.~L., {Stock}, J.~W., {et~al.} 2017, Astrobiology,
  17, 27

\bibitem[{Gillon {et~al.}(2017)Gillon, Triaud, Demory, Jehin, Agol, Deck,
  Lederer, de~Wit, Burdanov, Ingalls, Bolmont, Leconte, Raymond, Selsis,
  Turbet, Barkaoui, Burgasser, Burleigh, Carey, Chaushev, Copperwheat, Delrez,
  Fernandes, Holdsworth, Kotze, Van~Grootel, Almleaky, Benkhaldoun, Magain, \&
  Queloz}]{gillon2017}
Gillon, M., Triaud, A. H. M.~J., Demory, B.-O., {et~al.} 2017, Nature, 542, 456

\bibitem[{{Glasse} {et~al.}(2015){Glasse}, {Rieke}, {Bauwens},
  {Garc{\'\i}a-Mar{\'\i}n}, {Ressler}, {Rost}, {Tikkanen}, {Vandenbussche}, \&
  {Wright}}]{glasse2015}
{Glasse}, A., {Rieke}, G.~H., {Bauwens}, E., {et~al.} 2015, \pasp, 127, 686

\bibitem[{{Gravity Collaboration} {et~al.}(2019){Gravity Collaboration},
  {Lacour}, {Nowak}, {Wang}, {Pfuhl}, {Eisenhauer}, {Abuter}, {Amorim},
  {Anugu}, {Benisty}, {Berger}, {Beust}, {Blind}, {Bonnefoy}, {Bonnet},
  {Bourget}, {Brandner}, {Buron}, {Collin}, {Charnay}, {Chapron}, {Cl{\'e}net},
  {Coud{\'e} Du Foresto}, {de Zeeuw}, {Deen}, {Dembet}, {Dexter}, {Duvert},
  {Eckart}, {F{\"o}rster Schreiber}, {F{\'e}dou}, {Garcia}, {Garcia Lopez},
  {Gao}, {Gendron}, {Genzel}, {Gillessen}, {Gordo}, {Greenbaum}, {Habibi},
  {Haubois}, {Hau{\ss}mann}, {Henning}, {Hippler}, {Horrobin}, {Hubert},
  {Jimenez Rosales}, {Jocou}, {Kendrew}, {Kervella}, {Kolb}, {Lagrange},
  {Lapeyr{\`e}re}, {Le Bouquin}, {L{\'e}na}, {Lippa}, {Lenzen}, {Maire},
  {Molli{\`e}re}, {Ott}, {Paumard}, {Perraut}, {Perrin}, {Pueyo}, {Rabien},
  {Ram{\'{\i}}rez}, {Rau}, {Rodr{\'{\i}}guez-Coira}, {Rousset},
  {Sanchez-Bermudez}, {Scheithauer}, {Schuhler}, {Straub}, {Straubmeier},
  {Sturm}, {Tacconi}, {Vincent}, {van Dishoeck}, {von Fellenberg}, {Wank},
  {Waisberg}, {Widmann}, {Wieprecht}, {Wiest}, {Wiezorrek}, {Woillez},
  {Yazici}, {Ziegler}, \& {Zins}}]{Lacour2019}
{Gravity Collaboration}, {Lacour}, S., {Nowak}, M., {et~al.} 2019, \aap, 623,
  L11

\bibitem[{{Greene} {et~al.}(2016){Greene}, {Line}, {Montero}, {Fortney},
  {Lustig-Yaeger}, \& {Luther}}]{greene2016}
{Greene}, T.~P., {Line}, M.~R., {Montero}, C., {et~al.} 2016, \apj, 817, 17

\bibitem[{Gretzinger {et~al.}(2019)Gretzinger, Gross, Arriola, \&
  Withford}]{gretzinger2019}
Gretzinger, T., Gross, S., Arriola, A., \& Withford, M.~J. 2019, Opt. Express,
  27, 8626

\bibitem[{{Hanot} {et~al.}(2011){Hanot}, {Mennesson}, {Martin}, {Liewer},
  {Loya}, {Mawet}, {Riaud}, {Absil}, \& {Serabyn}}]{hanot2011}
{Hanot}, C., {Mennesson}, B., {Martin}, S., {et~al.} 2011, \apj, 729, 110

\bibitem[{Hedelt {et~al.}(2013)Hedelt, von Paris, Godolt, Gebauer, Grenfell,
  Rauer, Schreier, Selsis, \& Trautmann}]{hedelt2013}
Hedelt, P., von Paris, P., Godolt, M., {et~al.} 2013, Astronomy {\&}
  Astrophysics, 553, A9

\bibitem[{{Hsu} {et~al.}(2019){Hsu}, {Ford}, {Ragozzine}, \& {Ashby}}]{hsu2019}
{Hsu}, D.~C., {Ford}, E.~B., {Ragozzine}, D., \& {Ashby}, K. 2019, arXiv
  e-prints, arXiv:1902.01417

\bibitem[{{Janson}(2010)}]{janson2010}
{Janson}, M. 2010, \mnras, 408, 514

\bibitem[{Kaltenegger(2017)}]{kaltenegger2017}
Kaltenegger, L. 2017, Annual Review of Astronomy and Astrophysics, 55, 433

\bibitem[{{Kammerer} \& {Quanz}(2018)}]{kammerer2018}
{Kammerer}, J., \& {Quanz}, S.~P. 2018, \aap, 609, A4

\bibitem[{{Kitzmann} {et~al.}(2011){Kitzmann}, {Patzer}, {von Paris}, {Godolt},
  \& {Rauer}}]{kitzmann2011}
{Kitzmann}, D., {Patzer}, A.~B.~C., {von Paris}, P., {Godolt}, M., \& {Rauer},
  H. 2011, \aap, 531, A62

\bibitem[{{Kitzmann} {et~al.}(2010){Kitzmann}, {Patzer}, {von Paris}, {Godolt},
  {Stracke}, {Gebauer}, {Grenfell}, \& {Rauer}}]{kitzmann2010}
{Kitzmann}, D., {Patzer}, A.~B.~C., {von Paris}, P., {et~al.} 2010, \aap, 511,
  A66

\bibitem[{{Kopparapu} {et~al.}(2013){Kopparapu}, {Ramirez}, {Kasting}, {Eymet},
  {Robinson}, {Mahadevan}, {Terrien}, {Domagal-Goldman}, {Meadows}, \&
  {Deshpande}}]{kopparapu2013}
{Kopparapu}, R.~K., {Ramirez}, R., {Kasting}, J.~F., {et~al.} 2013, \apj, 765,
  131

\bibitem[{Kopparapu {et~al.}(2018)Kopparapu, H{\'e}brard, Belikov, Batalha,
  Mulders, Stark, Teal, Domagal-Goldman, \& Mandell}]{kopparapu2018}
Kopparapu, R.~k., H{\'e}brard, E., Belikov, R., {et~al.} 2018, The
  Astrophysical Journal, 856, 122

\bibitem[{{Kraus} {et~al.}(2016){Kraus}, {Ireland}, {Huber}, {Mann}, \&
  {Dupuy}}]{kraus2016}
{Kraus}, A.~L., {Ireland}, M.~J., {Huber}, D., {Mann}, A.~W., \& {Dupuy}, T.~J.
  2016, \aj, 152, 8

\bibitem[{{Kreidberg} \& {Loeb}(2016)}]{kreidberg2016}
{Kreidberg}, L., \& {Loeb}, A. 2016, \apjl, 832, L12

\bibitem[{{Kreidberg} {et~al.}(2014){Kreidberg}, {Bean}, {D{\'e}sert},
  {Benneke}, {Deming}, {Stevenson}, {Seager}, {Berta-Thompson}, {Seifahrt}, \&
  {Homeier}}]{kreidberg2014}
{Kreidberg}, L., {Bean}, J.~L., {D{\'e}sert}, J.-M., {et~al.} 2014, \nat, 505,
  69

\bibitem[{{Labadie} {et~al.}(2018){Labadie}, {Minardi}, {Mart{\'\i}n}, \&
  {Thomson}}]{labadie2018}
{Labadie}, L., {Minardi}, S., {Mart{\'\i}n}, G., \& {Thomson}, R.~R. 2018,
  Experimental Astronomy, 46, 433

\bibitem[{{Leconte} {et~al.}(2013){Leconte}, {Forget}, {Charnay}, {Wordsworth},
  {Selsis}, {Millour}, \& {Spiga}}]{leconte2013}
{Leconte}, J., {Forget}, F., {Charnay}, B., {et~al.} 2013, \aap, 554, A69

\bibitem[{{Leconte} {et~al.}(2015){Leconte}, {Forget}, \&
  {Lammer}}]{leconte2015}
{Leconte}, J., {Forget}, F., \& {Lammer}, H. 2015, Experimental Astronomy, 40,
  449

\bibitem[{L{\'e}ger {et~al.}(2019)L{\'e}ger, Defr{\`e}re, Garcia~Munoz, Godolt,
  Grenfell, Rauer, \& Tian}]{leger2019}
L{\'e}ger, A., Defr{\`e}re, D., Garcia~Munoz, A., {et~al.} 2019, Astrobiology,
  19, 797

\bibitem[{{Linz} {et~al.}(2019){Linz}, {Bhatia}, {Buinhas}, {Lezius}, {Ferrer},
  {F{\"o}rstner}, {Frankl}, {Philips-Blum}, {Steen}, {Bestmann}, {H{\"a}nsel},
  {Holzwarth}, {Krause}, \& {Pany}}]{linz2019}
{Linz}, H., {Bhatia}, D., {Buinhas}, L., {et~al.} 2019, arXiv e-prints,
  arXiv:1907.07989

\bibitem[{{L{\'o}pez-Morales} {et~al.}(2019){L{\'o}pez-Morales}, {Ben-Ami},
  {Gonzalez-Abad}, {Garc{\'\i}a-Mej{\'\i}a}, {Dietrich}, \&
  {Szentgyorgyi}}]{lopezmorales2019}
{L{\'o}pez-Morales}, M., {Ben-Ami}, S., {Gonzalez-Abad}, G., {et~al.} 2019,
  \aj, 158, 24

\bibitem[{{Lovis} {et~al.}(2017){Lovis}, {Snellen}, {Mouillet}, {Pepe},
  {Wildi}, {Astudillo-Defru}, {Beuzit}, {Bonfils}, {Cheetham}, \&
  {Conod}}]{lovis2017}
{Lovis}, C., {Snellen}, I., {Mouillet}, D., {et~al.} 2017, \aap, 599, A16

\bibitem[{Lustig-Yaeger {et~al.}(2018)Lustig-Yaeger, Meadows, Mendoza,
  Schwieterman, Fujii, Luger, \& Robinson}]{LustigYaeger2018}
Lustig-Yaeger, J., Meadows, V.~S., Mendoza, G.~T., {et~al.} 2018, The
  Astronomical Journal, 156, 301

\bibitem[{{Madhusudhan}(2018)}]{madhu2018}
{Madhusudhan}, N. 2018, Handbook of Exoplanets, ISBN 978-3-319-55332-0.
  Springer International Publishing AG, part of Springer Nature, 2018, id.104,
  104

\bibitem[{{Martin} {et~al.}(2012){Martin}, {Booth}, {Liewer}, {Raouf}, {Loya},
  \& {Tang}}]{martin2012}
{Martin}, S., {Booth}, A., {Liewer}, K., {et~al.} 2012, \ao, 51, 3907

\bibitem[{Mayor \& Queloz(1995)}]{mayor1995}
Mayor, M., \& Queloz, D. 1995, Nature, 378, 355

\bibitem[{Mayor {et~al.}(2011)Mayor, Marmier, Lovis, Udry, S{\'e}gransan, Pepe,
  Benz, Bertaux, Bouchy, Dumusque, Lo~Curto, Mordasini, Queloz, \&
  Santos}]{mayor2011}
Mayor, M., Marmier, M., Lovis, C., {et~al.} 2011, arXiv.org, arXiv:1109.2497

\bibitem[{{Meadows} {et~al.}(2018){Meadows}, {Reinhard}, {Arney}, {Parenteau},
  {Schwieterman}, {Domagal-Goldman}, {Lincowski}, {Stapelfeldt}, {Rauer},
  {DasSarma}, {Hegde}, {Narita}, {Deitrick}, {Lustig-Yaeger}, {Lyons},
  {Siegler}, \& {Grenfell}}]{meadows2018}
{Meadows}, V.~S., {Reinhard}, C.~T., {Arney}, G.~N., {et~al.} 2018,
  Astrobiology, 18, 630

\bibitem[{{Morley} {et~al.}(2017){Morley}, {Kreidberg}, {Rustamkulov},
  {Robinson}, \& {Fortney}}]{morley2017}
{Morley}, C.~V., {Kreidberg}, L., {Rustamkulov}, Z., {Robinson}, T., \&
  {Fortney}, J.~J. 2017, \apj, 850, 121

\bibitem[{{NASA}(2009)}]{null_depth}
{NASA}. 2009, {Exoplanet Interferometry Technology Milestone {\#}3 Report
  Broadband Starlight Suppression Demonstration}, Tech. rep.

\bibitem[{{Nielsen} {et~al.}(2019){Nielsen}, {De Rosa}, {Macintosh}, {Wang},
  {Ruffio}, {Chiang}, {Marley}, {Saumon}, {Savransky}, \&
  {Ammons}}]{nielsen2019}
{Nielsen}, E.~L., {De Rosa}, R.~J., {Macintosh}, B., {et~al.} 2019, \aj, 158,
  13

\bibitem[{Quanz {et~al.}(2015)Quanz, Crossfield, Meyer, Schmalzl, \&
  Held}]{quanz2015}
Quanz, S.~P., Crossfield, I., Meyer, M.~R., Schmalzl, E., \& Held, J. 2015,
  International Journal of Astrobiology, 14, 279

\bibitem[{{Quanz} {et~al.}(2018){Quanz}, {Kammerer}, {Defr{\`e}re}, {Absil},
  {Glauser}, \& {Kitzmann}}]{quanz2018}
{Quanz}, S.~P., {Kammerer}, J., {Defr{\`e}re}, D., {et~al.} 2018, in Society of
  Photo-Optical Instrumentation Engineers (SPIE) Conference Series, Vol. 10701,
  Optical and Infrared Interferometry and Imaging VI, 107011I

\bibitem[{{Rauer} {et~al.}(2014){Rauer}, {Catala}, {Aerts}, {Appourchaux},
  {Benz}, {Brandeker}, {Christensen-Dalsgaard}, {Deleuil}, {Gizon}, \&
  {Goupil}}]{rauer2014}
{Rauer}, H., {Catala}, C., {Aerts}, C., {et~al.} 2014, Experimental Astronomy,
  38, 249

\bibitem[{Rogers(2015)}]{rogers2015}
Rogers, L.~A. 2015, The Astrophysical Journal, 801, 41

\bibitem[{Sampaolo {et~al.}(2015)Sampaolo, Patimisco, Kriesel, Tittel,
  Scamarcio, \& Spagnolo}]{sampaolo2015}
Sampaolo, A., Patimisco, P., Kriesel, J.~M., {et~al.} 2015, Opt. Express, 23,
  195

\bibitem[{Schwieterman {et~al.}(2015)Schwieterman, Robinson, Meadows, Misra, \&
  Domagal-Goldman}]{schwieterman2015}
Schwieterman, E.~W., Robinson, T.~D., Meadows, V.~S., Misra, A., \&
  Domagal-Goldman, S. 2015, The Astrophysical Journal, 810, 57

\bibitem[{{Schwieterman} {et~al.}(2018){Schwieterman}, {Kiang}, {Parenteau},
  {Harman}, {DasSarma}, {Fisher}, {Arney}, {Hartnett}, {Reinhard}, {Olson},
  {Meadows}, {Cockell}, {Walker}, {Grenfell}, {Hegde}, {Rugheimer}, {Hu}, \&
  {Lyons}}]{schwieterman2018}
{Schwieterman}, E.~W., {Kiang}, N.~Y., {Parenteau}, M.~N., {et~al.} 2018,
  Astrobiology, 18, 663

\bibitem[{{Seager} {et~al.}(2013){Seager}, {Bains}, \& {Hu}}]{seager2013}
{Seager}, S., {Bains}, W., \& {Hu}, R. 2013, \apj, 777, 95

\bibitem[{Seager \& Deming(2010)}]{seager2010}
Seager, S., \& Deming, D. 2010, Annual Review of Astronomy and Astrophysics,
  48, 631

\bibitem[{Seager {et~al.}(2005)Seager, Turner, Schafer, \& Ford}]{seager2005}
Seager, S., Turner, E.~L., Schafer, J., \& Ford, E.~B. 2005, Astrobiology, 5,
  372

\bibitem[{Sing {et~al.}(2016)Sing, Fortney, Nikolov, Wakeford, Kataria, Evans,
  Aigrain, Ballester, Burrows, Deming, D{\'e}sert, Gibson, Henry, Huitson,
  Knutson, des Etangs, Pont, Showman, Vidal-Madjar, Williamson, \&
  Wilson}]{sing2016}
Sing, D.~K., Fortney, J.~J., Nikolov, N., {et~al.} 2016, Nature, 529, 59

\bibitem[{{Snellen} {et~al.}(2015){Snellen}, {de Kok}, {Birkby}, {Brandl},
  {Brogi}, {Keller}, {Kenworthy}, {Schwarz}, \& {Stuik}}]{snellen2015}
{Snellen}, I., {de Kok}, R., {Birkby}, J.~L., {et~al.} 2015, \aap, 576, A59

\bibitem[{{Sozzetti} \& {de Bruijne}(2018)}]{sozzetti2018}
{Sozzetti}, A., \& {de Bruijne}, J. 2018, Handbook of Exoplanets, ISBN
  978-3-319-55332-0. Springer International Publishing AG, part of Springer
  Nature, 2018, id.81, 81

\bibitem[{{Stark} {et~al.}(2014){Stark}, {Roberge}, {Mandell}, \&
  {Robinson}}]{stark2014}
{Stark}, C.~C., {Roberge}, A., {Mandell}, A., \& {Robinson}, T.~D. 2014, \apj,
  795, 122

\bibitem[{Takahashi {et~al.}(2017)Takahashi, Enya, Haze, Kataza, Kotani,
  Matsuhara, Kamiya, Yamamuro, Bierden, Cornelissen, Lam, \&
  Feinberg}]{takahashi2017}
Takahashi, A., Enya, K., Haze, K., {et~al.} 2017, Appl. Opt., 56, 6694

\bibitem[{{Tepper} {et~al.}(2017){Tepper}, {Labadie}, {Diener}, {Minardi},
  {Pott}, {Thomson}, \& {Nolte}}]{tepper2017}
{Tepper}, J., {Labadie}, L., {Diener}, R., {et~al.} 2017, \aap, 602, A66

\bibitem[{{Tinetti} {et~al.}(2018){Tinetti}, {Drossart}, {Eccleston},
  {Hartogh}, {Heske}, {Leconte}, {Micela}, {Ollivier}, {Pilbratt}, \&
  {Puig}}]{tinetti2018}
{Tinetti}, G., {Drossart}, P., {Eccleston}, P., {et~al.} 2018, Experimental
  Astronomy, 46, 135

\bibitem[{von Paris {et~al.}(2013)von Paris, Hedelt, Selsis, Schreier, \&
  Trautmann}]{vonparis2013}
von Paris, P., Hedelt, P., Selsis, F., Schreier, F., \& Trautmann, T. 2013,
  Astronomy {\&} Astrophysics, 551, A120

\bibitem[{{Weiss} {et~al.}(2018){Weiss}, {Marcy}, {Petigura}, {Fulton},
  {Howard}, {Winn}, {Isaacson}, {Morton}, {Hirsch}, {Sinukoff}, {Cumming},
  {Hebb}, \& {Cargile}}]{weiss2018}
{Weiss}, L.~M., {Marcy}, G.~W., {Petigura}, E.~A., {et~al.} 2018, \aj, 155, 48

\bibitem[{Wolfgang {et~al.}(2016)Wolfgang, Rogers, \& Ford}]{wolfgang2016}
Wolfgang, A., Rogers, L.~A., \& Ford, E.~B. 2016, The Astrophysical Journal,
  825, 19

\bibitem[{Yu {et~al.}(2012)Yu, Wadsworth, \& Knight}]{yu2012}
Yu, F., Wadsworth, W.~J., \& Knight, J.~C. 2012, Opt. Express, 20, 11153

\bibitem[{{Zamkotsian} {et~al.}(2017){Zamkotsian}, {Lanzoni}, {Barette},
  {Grassi}, {Vors}, {Helmbrecht}, {Marchis}, \& {Teichman}}]{zamkotsian2017}
{Zamkotsian}, F., {Lanzoni}, P., {Barette}, R., {et~al.} 2017, in Society of
  Photo-Optical Instrumentation Engineers (SPIE) Conference Series, Vol. 10116,
  SPIE, 101160M

\bibitem[{{Zsom} {et~al.}(2013){Zsom}, {Seager}, {de Wit}, \&
  {Stamenkovi{\'c}}}]{zsom2013}
{Zsom}, A., {Seager}, S., {de Wit}, J., \& {Stamenkovi{\'c}}, V. 2013, \apj,
  778, 109

\end{thebibliography}

\pagebreak
\section*{Team of co-authors and supporters (in alphabetic order)}
\begin{table}[!h]
    \centering
    \footnotesize
\begin{tabular}{ll|l|l|l}
& Name & Affiliation & Country & Email \\\hline\hline

& Olivier Absil  & University of Li\`ege & Belgium & olivier.absil@uliege.be \\
 
 &Daniel Angerhausen & University of Bern & Switzerland & daniel.angerhausen@csh.unibe.ch \\

 &Willy Benz & University of Bern & Switzerland & willy.benz@space.unibe.ch \\

 &Xavier Bonfils & Universit\'e Grenoble Alpes & France & xavier.bonfils@univ-grenoble-alpes.fr \\

 &Jean-Philippe Berger & Universit\'e Grenoble Alpes & France & bergejea@univ-grenoble-alpes.fr \\
 
 &Matteo Brogi & Univ. of Warwick & UK & m.brogi@warwick.ac.uk \\

 &Juan Cabrera & German Aerospace Center (DLR)  & Germany & Juan.Cabrera@dlr.de \\

 &William C. Danchi & Goddard Space Flight Center  & USA & william.c.danchi@nasa.gov \\

 &Denis Defr\`ere  & Li\`ege Space Center & Belgium & ddefrere@uliege.be \\
 
  &Ewine van Dishoeck & Leiden Observatory & Netherlands & ewine@strw.leidenuniv.nl \\

 &David Ehrenreich  & University of Geneva & Switzerland & david.ehrenreich@unige.ch \\

 &Steve Ertel  & LBT Observatory & USA & sertel@email.arizona.edu \\

 &Jonathan Fortney & UC Santa Cruz & USA & jfortney@ucsc.edu \\

 &Scott Gaudi & Ohio State University & USA & gaudi.1@osu.edu \\

  &Julien Girard & Space Telescope Science Institute  & USA & jgirard@stsci.edu \\

  &Adrian Glauser & ETH Zurich  & Switzerland & glauser@phys.ethz.ch \\

  &John Lee Grenfell & German Aerospace Center  (DLR) & Germany & Lee.Grenfell@dlr.de \\

 &Michael Ireland & Australian National University & Australia & michael.ireland@anu.edu.au \\

 &Markus Janson & University of Stockholm & Sweden & markus.janson@astro.su.se \\

  & Jens Kammerer & Australian National University & Australia & Jens.Kammerer@anu.edu.au \\

  &Daniel Kitzmann & University of Bern & Switzerland & daniel.kitzmann@csh.unibe.ch \\
  
 &Stefan Kraus & University of Exeter & UK & s.kraus@exeter.ac.uk \\ 
 
 &Oliver Krause & Max Planck Inst. for Astronomy & Germany & krause@mpia.de \\
 
 &Lucas Labadie & University of Cologne & Germany & labadie@ph1.uni-koeln.de \\
 
  &Sylvestre Lacour & Observatoire de Paris & France & sylvestre.lacour@obspm.fr \\
 
 &Tim Lichtenberg & University of Oxford & UK & tim.lichtenberg@physics.ox.ac.uk \\
 
 &Michael Line & Arizona State University & USA & mrline@asu.edu \\

 &Hendrik Linz & Max Planck Inst. for Astronomy & Germany & linz@mpia.de \\

 &Jérôme Loicq & Li\`ege Space Center & Belgium & j.loicq@uliege.be \\

 &Bertrand Mennesson & Jet Propulsion Laboratory & USA & bertrand.mennesson@jpl.nasa.gov \\
 
 &Michael Meyer & University of Michigan & USA & mrmeyer@umich.edu \\

 &Yamila Miguel & Leiden Observatory & Netherlands & ymiguel@strw.leidenuniv.nl \\

  &John Monnier & University of Michigan & USA & monnier@umich.edu \\

& Mamadou N’Diaye & Observatoire de la C\^ote d'Azur & France & mamadou.ndiaye@oca.eu \\

 &Enric Pall\'e & Instituto de Astrofisica de Canarias & Spain & epalle@iac.es \\

 &Didier Queloz & University of Cambridge & UK & dq212@cam.ac.uk \\
 
  &Heike Rauer & German Aerospace Center (DLR) & Germany & Heike.Rauer@dlr.de \\
  
  &Ignasi Ribas & Institut de Ciències de l'Espai  & Spain & iribas@ice.cat \\
  
 &Sarah Rugheimer & University of Oxford & UK & sarah.rugheimer@physics.ox.ac.uk \\

  &   Franck Selsis & Lab. d'astrophysique de Bordeaux & France & franck.selsis@u-bordeaux.fr \\

  &   Gene Serabyn & Jet Propulsion Laboratory, Caltech & USA & gene.serabyn@jpl.nasa.gov \\

  &   Ignas Snellen & Leiden Observatory & Netherlands & snellen@strw.leidenuniv.nl \\
  
  &   Alessandro Sozzetti & INAF - Torino & Italy &  alessandro.sozzetti@inaf.it \\
 
   &   Karl R. Stapelfeldt & Jet Propulsion Laboratory, Caltech & USA & Karl.R.Stapelfeldt@jpl.nasa.gov \\
 
   &   Amaury Triaud & University of Birmingham & UK &  A.Triaud@bham.ac.uk \\
 
 &   St\'ephane Udry & University of Geneva & Switzerland & Stephane.Udry@unige.ch \\
 
 &   Mark Wyatt & University of Cambridge & UK & wyatt@ast.cam.ac.uk \\
\end{tabular}
    \label{tab:my_label}
\end{table}

\end{document}